\documentclass[11pt,a4paper]{article} 
\usepackage[utf8]{inputenc}
\usepackage{amsmath,amsthm}
\usepackage[unicode,colorlinks=true,linkcolor=blue,citecolor=blue]{hyperref}
\usepackage[all]{hypcap}
\usepackage{textcomp}
\usepackage{amsfonts}
\usepackage{mathrsfs}
\usepackage{lettrine}
\usepackage{amssymb}
\usepackage{graphicx}
\usepackage{graphicx}
\usepackage{vmargin}
\usepackage{eurosym}
\setmarginsrb{18mm}{15mm}{18mm}{10mm}{10pt}{0mm}{0pt}{20mm}
\makeindex
\usepackage{shapepar}
\usepackage{array}
\usepackage{bbm}
\usepackage{multicol}
\usepackage{bm}
\usepackage[superscript]{cite}
\usepackage{lipsum}
\usepackage{dcolumn}
\usepackage{braket}
\usepackage{stmaryrd}
\usepackage{color}
\usepackage{rotating}
\usepackage{longtable}
\usepackage{tocloft}
\DeclareMathAlphabet{\mathpzc}{OT1}{pzc}{m}{it}
\usepackage{tikz}
\usepackage{colortbl}
\usepackage{multimedia}
\usepackage{braket}
\usepackage{amsfonts}
\usepackage{xcolor}
\definecolor{olive}{rgb}{0.3, 0.4, .1}
\definecolor{fore}{RGB}{249,242,215}
\definecolor{back}{RGB}{51,51,51}
\definecolor{title}{RGB}{255,0,90}
\definecolor{BlueViolet}{RGB}{138,43,226}
\definecolor{dgreen}{rgb}{0.,0.6,0.}
\definecolor{gold}{rgb}{1.,0.84,0.}
\definecolor{JungleGreen}{cmyk}{0.99,0,0.52,0}
\definecolor{BlueGreen}{cmyk}{0.85,0,0.33,0}
\definecolor{RawSienna}{cmyk}{0,0.72,1,0.45}
\definecolor{Magenta}{cmyk}{0,1,0,0}
\definecolor{wood}{RGB}{139,115,85}
\definecolor{dorange}{RGB}{255,127,0}
\definecolor{dolive}{RGB}{85,107,47}
\definecolor{drg}{RGB}{255,165,0}
\def\checkmark{\tikz\fill[scale=0.4](0,.35) -- (.25,0) -- (1,.7) -- (.25,.15) -- cycle;} 
\usepackage{arydshln}
\usepackage{multirow}

\newtheorem{theorem}{Theorem}[section]
\newtheorem{corollary}{Corollary}[theorem]
\newtheorem{lemma}[theorem]{Lemma}

\usepackage{simplewick}

\newcommand\tvecpre{{\mathpalette\raiseT\top}}
\newcommand\raiseT[2]{\raisebox{-0.4ex}{$#1#2$}}
\newcommand\tvec{\mkern-2mu^\tvecpre\!}
\newcommand{\tmat}{\mkern-2mu ^\top\!}

\begin{document}

\title{\vspace*{-2cm}  A comprehensive, self-contained derivation of the one-body density matrices from single-reference excited-state calculation methods using the equation-of-motion formalism}

\date{}

\maketitle

\vspace*{-1.4cm}

\noindent \begin{center}
Thibaud Etienne$^{\dag,*}$
\end{center}

$\;$

\noindent $^\dag$ ICGM, University of Montpellier, CNRS, ENSCM, Montpellier, France

\noindent $^*$ thibaud.etienne@umontpellier.fr

$$$$


\noindent \textbf{Abstract} In this contribution we review in a rigorous, yet comprehensive fashion the assessment of the one-body reduced density matrices derived from the most used single-reference excited-state calculation methods in the framework of the equation-of-motion formalism. Those methods are separated into two types: those which involve the coupling of a deexcitation operator to a single-excitation transition operator, and those which do not involve such a coupling. The case of many-body auxiliary wave functions for excited states is also addressed. For each of these approaches we were interested in deriving the elements of the one-body transition and difference density matrices, and to highlight their particular structure. This has been accomplished by applying a decomposition of integrals involving one-determinant quantum electronic states on which two or three pairs of second quantization operators can act. Such a decomposition has been done according to a corollary to Wick's theorem, which is brought in a comprehensive and detailed manner. A comment is also given about the consequences of using the equation-of-motion formulation in this context, and the two types of excited-state calculation methods (with and without coupling excitations to deexcitations) are finally compared from the point of view of the structure of their transition and difference density matrices.

$\;$


\newpage

\section{Introduction}

Amongst the physical processes studied by theoreticians, light-induced molecular electronic transitions, due to their potential involvement in the development of new technologies, are now involving careful consideration from both experimental and theoretical (academic and industrial) research communities. Ubiquitous in biology and chemistry, they are under active study in the context of many crucial societal issues of our time. These issues include, among others, public health with optogenetics\cite{dieter_optogenetics_2019} and phototherapy\cite{youngsik_therapy_2019}, or climate change with the design of efficient photovoltaic devices\cite{editorial_perovskite_2019}. Quantum-molecular science, and in particular quantum-chemical excited-state calculation methods, are now established as a reliable mean to gain fundamental qualitative and quantitative insights into light-induced physical or chemical processes at the atomic level. Indeed, the theoretical characterization of electronic excited states is the aim of an increasing number of scientific contributions, and the tools used for such analyses involve generally the objects originating directly from the calculation of the excited states.\cite{dreuw_single-reference_2005,plasser_new_2014,savarese_metrics_2017} Depending on the method used for computing the electronic transitions, the objects derived from these calculations will not have the same structure and the same properties.\cite{luzanov_electron_2010} Since the analysis of the nature of the excited states generally relies on the use of these objects (in particular the transition and difference density matrices, that will both be at the center of this contribution), either from a qualitative point of view using exciton analysis\cite{luzanov_interpretation_1980,furche_density_2001,tretiak_density_2002,
martin_natural_2003,batista_et_martin_natural_2004,tretiak_exciton_2005,dreuw_single-reference_2005,
mayer_using_2007,surjan_natural_2007,wu_exciton_2008,luzanov_electron_2010,li_time-dependent_2011,plasser_analysis_2012,plasser_new_2014,pluhar_visualizing_2018,plasser_newbis_2014,bappler_exciton_2014,
mewes2015communication,li_particlehole_2015,li_particlehole_2016,etienne2015transition,plasser2015statistical,
poidevin_truncated_2016,wenzel_physical_2016,
plasser2016entanglement,savarese_metrics_2017,plasser_detailed_2017,
mai_quantitative_2018,mewes_benchmarking_2018,etienne_theoretical_2018,
skomorowski_real_2018,park_low_2018} or one-particle charge density functions and their corresponding density matrices,\cite{HeadGordonJPhysChem1995,dreuw_single-reference_2005,plasser_new_2014,plasser_newbis_2014,ronca_charge-displacement_2014,ronca_density_2014,etienne_new_2014,
pastore2017unveiling} or under a quantitative perspective using descriptors,\cite{luzanov_interpretation_1980,peach_excitation_2008,luzanov_electron_2010,
le_bahers_qualitative_2011,
garcia_evaluating_2013,GuidoJChemTheoryComput2013,etienne_toward_2014,
etienne_new_2014,guido_effective_2014,etienne_probing_2015,plasser2015statistical,
wenzel_physical_2016,poidevin_truncated_2016,plasser2016entanglement,savarese_metrics_2017,pastore2017unveiling,
plasser_detailed_2017,mai_quantitative_2018,etienne_theoretical_2018,
barca_excitation_2018,mewes_benchmarking_2018,
campetella_quantifying_2019} a proper knowledge of their structure is required for selecting the right post-processing strategy. Unfortunately, while the structure of the objects derived from the calculations are often known for the most common calculation methods, in most of the cases this structure is given without a detailed derivation. While the derivation itself might seem long and complicated, in this tutorial we show how we can use a simple formulation of the problem and find rapid routes for assessing the particular objects we target: the one-body reduced density matrices\cite{cioslowski_many-electron_2000,coleman_structure_1963,davidson_reduced_1976,lowdin_quantum_1955,mcweeny_recent_1960,
mcweeny_methods_1992,mcweeny_density_1959,
mcweeny_density_1961} (1-RDM), about which a brief reminder is provided in the text. In this contribution we start by revisiting a known and used tool for assessing integrals involving quantum electronic states on which pairs of second quantization operators have acted, i.e., a variant of Wick's theorem.\cite{wick_evaluation_1950,gross_many-particle_1991,kutzelnigg_normal_1997,surjan_second_1989,shavitt_many-body_2009,wilson_methods_1992} Since the application of this theorem can be somehow tedious, we provide here a diagrammatic, yet rigorous, derivation of the integrals one can assess using this tool. 

This theorem is then used in the context of the assessment of transition and difference density matrix elements for characterizing excited states originating from equation-of-motion\cite{
rowe_equations--motion_1968,
yeager_equations_1975,
mccurdy_equations_1977,
lynch_excited_1982} (EOM) calculations performed using two types of methods.

The first type of methods involves a single-excitation transition operator, added to a deexcitation one. The basic principles of the EOM formalism are briefly recalled in first quantization, before the transition operator for the first class of methods mentioned above is introduced into this framework. The methods from this first class, i.e., the Random Phase Approximation\cite{ehrenreich_self-consistent_1959,
rowe_equations--motion_1968,
yeager_equations_1975,
mccurdy_equations_1977,
mckoy_equations_1980,
joergensen_second_1981,
lynch_excited_1982,
schirmer_new_1996,
furche_developing_2008,
toulouse_range-separated_2010,
sauer_molecular_2011,
chatterjee_excitation_2012,
pernal_intergeminal_2014,
pastorczak_correlation_2018,
pernal_correlation_2018,
pernal_electron_2018} (RPA), the Time-Dependent Hartree-Fock theory\cite{
mclachlan_time-dependent_1964,
rowe_equations--motion_1968,
dalgaard_timedependent_1980,
joergensen_second_1981,
lynch_excited_1982,
mcweeny_time-dependent_1983,
yeager_generalizations_1984,
hirata_configuration_1999,
dreuw_single-reference_2005,
crespo-otero_recent_2018
} (TDHF), the Time-Dependent Density Functional Theory\cite{
casida_time-dependent_1995,
jamorski_dynamic_1996,
casida_time-dependent_1996,
casida_molecular_1998,
stratmann_efficient_1998,
hirata_configuration_1999,
hirata_time-dependent_1999,
furche_density_2001,
furche_adiabatic_2002,
shao_spinflip_2003,
furche_erratum:_2004,
furche_iii_2005,
dreuw_single-reference_2005,
pernal_time-dependent_2007,
elliott_excited_2009,
ipatov_excited-state_2009,
casida_time-dependent_2009,
sauer_molecular_2011,
ullrich_time-dependent_2011,
ferre_density-functional_2016,
park_low_2018,
crespo-otero_recent_2018
} (TDDFT) and the Bethe-Salpeter Equation\cite{
rocca_ab_2010,
jacquemin_is_2017,
blase_bethesalpeter_2018,
gui_accuracy_2018,
krause_implementation_2018,
leng_gw_2018
} (BSE), have a central equation with the same structure, and the objects we will use in this contribution have the same physical interpretation within each method. 

If the single excitations are not coupled to deexcitations, as in the Configuration Interaction Singles\cite{
foresman_toward_1992,
hirata_configuration_1999,
dreuw_single-reference_2005,
sauer_molecular_2011,
crespo-otero_recent_2018
} (CIS) or the Tamm-Dancoff Approximation\cite{
hirata_time-dependent_1999,
hirata_configuration_1999,
dreuw_single-reference_2005,
sauer_molecular_2011,
chantzis_is_2013,
crespo-otero_recent_2018
} (TDA), we have another class of methods, with objects having different structure and properties. These methods will be discussed in this paper. Note also that the Algebraic Diagrammatic Construction\cite{
mertins_algebraic_1996,
schirmer_intermediate_2004,
dreuw_algebraic_2015,
wenzel_physical_2016,
crespo-otero_recent_2018
} (ADC) involves matrices with similar interpretation. They are constructed in the space of the so-called ``intermediate states"\cite{mertins_algebraic_1996,schirmer_intermediate_2004} and lead to density matrices in the intermediate states space with the same structure as the CIS and TDA ones in the space of singly-excited Slater determinants. Additionally, the Linear Response (LR) variant of the single-reference Coupled Cluster\cite{
dalgaard_aspects_1983,
koch_coupled_1990,
koch_excitation_1990,
christiansen_second-order_1995,
hattig_cc2_2000,
hattig_implementation_2002,
hattig_structure_2005,
sauer_molecular_2011,
sneskov_excited_2012,
helmich_pair_2013,
crespo-otero_recent_2018
} (CC) theory for excited states can be used for producing matrices that can be compared to those from CIS, TDA and ADC, and classified among the second type of methods investigated in this contribution. Indeed, as for ADC and TD-DFRT\cite{casida_time-dependent_1995} (where ``R" stands for ``Response"), in the LR-CC framework an Auxiliary Many-Body Wave function (AMBW) has also been proposed for the assignment of excited states ansätze.\cite{crespo-otero_recent_2018} For ADC, this has been done in addition to the ansatz already introduced in the intermediate states basis. The three AMBWs are introduced as a linear combination of singly-excited Slater determinants so that, after a necessary renormalization,\cite{etienne_theoretical_2018} these ansätze can produce transition and difference density matrices that will belong to the second class of matrices we will derive in this tutorial, though the working equations for TD-DFRT (LR-CC) include (un)coupled excitations and deexcitations contributions to the transition energy. They are also mentioned in this report since, according to the results presented below, those ansätze for excited states can be formally considered, combined with the single-reference ground electronic state, as a compatible approximation to the EOM transition operator for producing the matrices of interest to characterize the electronic transition, in the same way the CIS and TDA ansätze do.

The two types of methods (including or not the deexcitation operator) mentioned above will finally be compared, based on the structure of their respective 1-RDMs, but also on the interpretation of the nature of their generating operator, and the consequence of its use in the EOM formalism.

\section{Hypotheses and theoretical background}

Our $N$-electron reference ground state wave function values $\psi _0 (\textbf{s}_1, \dotsc , \textbf{s}_N)$ are single Slater determinants written in the orthonormal, local basis of $L$ real-valued one-particle wave functions called spinorbitals $\left( \varphi _r \right) _{r=1}^L$ ($N$ of them being singly occupied, and $L-N$ being unoccupied and called ``virtual") where, since electrons are indiscernible, ${\textbf{s}}_i$ represents the four \textbf{s}pin-\textbf{s}patial coordinates of electron $i$, with $1\leq i \leq N$. 

For a given quantum electronic state ${\ket{\psi _n}}$ ($n \geq 0$), we have $\psi  _n(\textbf{s}_1, \dotsc , \textbf{s}_N) = \braket{\textbf{s}_1, \dotsc , \textbf{s}_N |\psi _n}$. To any $\ket{\psi _n}$ corresponds a 1--RDM written $\bm{\gamma} _n$ in the spinorbital basis. We therefore start by recalling the expression of the $\bm{\gamma}_n$ elements for any $\ket{\psi _n}$, as well as the expression of the transition density matrix ($\bm{\gamma}^{0\rightarrow n}$) elements corresponding to the transition from the ground state ($n = 0$) to any excited state ($n \neq 0$). In second quantization (see Appendix \ref{app:SQ}), those matrix elements read
\begin{equation}
(\bm{\gamma}_n)_{r,s} = \braket{\psi _n  | \hat{r}^\dag \hat{s}| \psi _n }, \quad (\bm{\gamma}^{0\rightarrow n})_{s,r} = \braket{\psi _0 | \hat{r}^\dag \hat{s}| \psi _n}, \label{eq:dm_and_tdms}
\end{equation}
with $r$ and $s$ ranging from 1 to $L$. Any $\hat{q}$ operator (replace here $q$ by $r$ or $s$ for instance) is the annihilation operator when it is acting on a ket placed on its right, and a creation operator when it is acting on a bra placed on its left. On the other hand, any $\hat{q}^\dag$ operator is the creation operator when it is acting on a ket placed on its right, and an annihilation operator when it is acting on a bra placed on its left.

Note that the formalism reported in this contribution is presented for spinorbitals, and could be generalized to single-particle orbitals which would be mixtures of spin-up and spin-down states. Such situations arise in noncollinear spin systems, and they are of interest in the so-called spin-flip approach\cite{shao_spinflip_2003} to TDDFT or TDHF. 

\subsection{Integrals of products of pairs of fermionic creation/annihilation operators}

In this section, no restriction has been imposed regarding the attribution of the spinorbitals pointed by the second quantization operators to a given space (e.g., occupied or virtual canonical subspace).

In this contribution we will mostly be dealing with integrals containing products of second quantization operators, such as
\begin{equation}
\mathscr{I}_M = \braket{\psi _0 | \underbrace{\hat{u}^\dag \hat{v}}_{k=1}\underbrace{\hat{w}^\dag \hat{x}}_{k=2} \cdots \underbrace{\hat{y}^\dag \hat{z}}_{k=M}| \psi _0} = \left\langle \psi _0 \left| \prod _{k = 1}^M {}^{[k]}\hat{c} _1 {}^{[k]}\hat{c}_{2} \right| \psi _0\right\rangle \label{eq:I_M}
\end{equation}
with, in this example, ${}^{[k]}\hat{c} _1$ (respectively, ${}^{[k]}\hat{c} _2$) being always a creation (respectively, annihilation) operator when acting on a given ket:
\begin{equation}
{}^{[1]}\hat{c} _1 {}^{[1]}\hat{c}_{2} = \hat{u}^\dag \hat{v}, \quad {}^{[2]}\hat{c} _1 {}^{[2]}\hat{c}_{2} = \hat{w}^\dag \hat{x}, \quad {}^{[M]}\hat{c} _1 {}^{[M]}\hat{c}_{2} = \hat{y}^\dag \hat{z},
\end{equation}
but first we would like to recall that the expectation value of a chain of $M$ pairs of second quantization operators (i.e., a $Q$-operator, with $Q = 2M$) can be decomposed into a sum of products of two-operator integrals. Since we are dealing with fermionic second quantization operators, a $(-1)^{m_h}$ sign is attributed to each product entering such a sum and the integral can be assessed as
\begin{equation}
\left\langle \psi _0 \left |\, \prod _{k=1}^{2M} \hat{Q}_k \, \right|\psi _0\right\rangle = \sum _{h = 1} ^{(2M-1)!!} (-1)^{m_h} \underbrace{\prod _{e=1}^{M}  {\left\langle\psi _0 \left| {}^{[e]}\hat{q} _{1,h} {}^{[e]}\hat{q}_{2,h} \right| \psi _0\right\rangle}}_{\displaystyle \mathscr{P}_h} \label{eq:initial_I}
\end{equation}
where the $\hat{Q}_k$ and the ${}^{[e]}\hat{q} _{i,h}$ operators can be either creation or annihilation operators. This result is a corollary to Wick's theorem, extensively discussed and demonstrated in Appendix \ref{app:wick}.

The number $\mathscr{N}_Q$ of possible products of integrals for a given $Q$-operator, is
\begin{equation}
\mathscr{N}_Q = (2M-1)!! = \dfrac{(2M)!}{M! \, 2^M} .
\end{equation}
The details about the deduction of $\mathscr{N}_Q$ are given in Appendix \ref{app:NQ}.

Note that these results are not restricted to integrals with products of pairs of creation/annihilation operators as in integral $\mathscr{I}_M$ from equation \eqref{eq:I_M}; they are relative to integrals with products of any second quantization operators, given that the total number $Q$ of these operators is even. 

In the particular case of integrals such as $\mathscr{I}_M$ in equation \eqref{eq:I_M}, that will be of interest to us in this paper, we see that due to the structure of its $Q$-operator the expectation value can be expanded into a sum of two-operator integrals as in Eq. \eqref{eq:initial_I} and reduced to the sum of its $M!$ non-vanishing terms
\begin{equation}
\mathscr{I}_M = \sum _{\ell = 1} ^{M!} (-1)^{f_\ell} \underbrace{ \prod _{p=1}^{M}{\left\langle\psi _0 \left| {}^{[p]}\hat{t} _{1,\ell} \;{}^{[p]}\hat{t}_{2,\ell} \right| \psi _0\right\rangle}}_{\displaystyle{\mathscr{P}_\ell}}  \label{eq:final_I}
\end{equation}
where again, ${}^{[p]}\hat{t} _{1,\ell}$ and ${}^{[p]}\hat{t} _{2,\ell}$ are creation and annihilation operators, respectively. We will see later how the $f _\ell$ value can be determined. 
\subsubsection{Application to four- and six-operator integrals}
In this section we will expose how the general formalism formulated above can be applied to the decomposition of integrals with two and three pairs of second quantization operators. These results will then be applied to the computation of matrix elements derived from excited-state calculation methods of different types.

$\;$

\noindent \textbf{\textcolor{black}{Four-operator integrals}}

$\;$

\begin{figure}[h!]
\begin{center}
\includegraphics[scale=1]{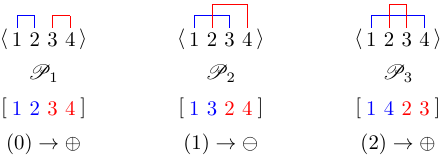}
\caption{Possible decomposition of integrals with two ordered pairs of creation/annihilation second quantization operators into the sum of three products of integrals. The number between parentheses is the number $m_h$ of times a pairing line crosses others (the number of times a line of a given color crosses lines with different colors) and the sign on the right of the arrow is the $\displaystyle (-1)^{m_h}$ signature. The reordered labels of the operators are reported between square brackets.}\label{fig:Pa_to_Pc}
\end{center}
\end{figure}

\noindent We start by writing any four-operator integral as
\begin{equation}
\mathscr{I}_2 = \left\langle \psi _0 \left| \hat{u}^\dag \hat{v} \hat{w}^\dag \hat{x} \right| \psi _0\right\rangle \label{eq:I2}
\end{equation}
The $u$, $v$, $w$, and $x$ were randomly chosen and point any spinorbital (occupied or virtual) for the moment. We put labels on the second quantization operators:
\begin{equation}
\hat{u}^\dag \rightarrow 1 , \quad \hat{v} \rightarrow 2 , \quad \hat{w}^\dag \rightarrow 3 , \quad \hat{x} \rightarrow 4,
\end{equation}
and we schematically write
\begin{equation}
\left\langle \psi _0 \left| \hat{u}^\dag \hat{v} \hat{w}^\dag \hat{x} \right| \psi _0\right\rangle = \braket{\,1\,2\,3\,4\,}.
\end{equation}
The decomposition \eqref{eq:initial_I} is illustrated for this case in figure \ref{fig:Pa_to_Pc}, where we highlighted the pairing of operators with colored lines. From the left to the right, if a pairing line arises before another one is closed (as in $\mathscr{P}_2$ for instance), it has to override the unterminated pairing line coming from its left. The sequence between square brackets is the new sequence of operators following the pairing chronological order from the left to the right. The number between parentheses is the number of times a colored line crosses other lines of different colors ($m_h$ in Eq. \eqref{eq:initial_I}). For instance, in $\mathscr{P}_2$, the contraction in red crosses once the blue one, while in $\mathscr{P}_3$ it crosses it twice (once when rising up, once when going down to 3). This number also corresponds to the number of times that numbers are superior to other(s) after themselves in the rearranged sequence. For example, in the sequence (not existing in the decomposition presented in figure \ref{fig:Pa_to_Pc} since the pairing is done from the left to the right) $\braket{ \,4 \,3 \,1 \,2\,}$, 4 is superior to 3, 1 and 2, while 3 is superior to 1 and 2, so in total the number $m_h$ of such an unexisting decomposition would be 5.

$\;$

\vspace*{-0.4cm}
\noindent In a general four-operator integral decomposition, we have, according to Eq. \eqref{eq:initial_I},
\begin{equation}
\mathscr{I}_2 = (-1)^{m_1}\mathscr{P}_\mathrm{1} + (-1)^{m_2}\mathscr{P}_2 + (-1)^{m_3}\mathscr{P}_3 \label{eq:I2_decomposition}
\end{equation}
with $m_1 = 0$, $m_2 = 1$, $m_3 = 2$. The case of equation \eqref{eq:I2} for example, reads
\begin{equation}
\mathscr{P}_\mathrm{1} = \braket{\psi _0 |\!\! \stackrel{1}{\hat{u}} \left.\!\!^\dag\right. \!\! \stackrel{2}{\hat{v}} \!\!| \psi _0}\braket{\psi _0 |\!\! \stackrel{3}{\hat{w}} \left.\!\!^\dag\right. \!\! \stackrel{4}{\hat{x}} \!\!| \psi _0} 
, \; 
\mathscr{P}_\mathrm{2} = \braket{\psi _0 |\!\! \stackrel{1}{\hat{u}} \left.\!\!^\dag\right. \!\! \stackrel{3}{\hat{w}} \left. \!\! ^\dag \right. \!\!| \psi _0}\braket{\psi _0 |\!\! \stackrel{2}{\hat{v}} \left.\!\! \right. \!\! \stackrel{4}{\hat{x}} \left. \!\!   \right. \!\!| \psi _0} 
, \; 
\mathscr{P}_\mathrm{3} = \braket{\psi _0 |\!\! \stackrel{1}{\hat{u}} \left.\!\!^\dag\right. \!\! \stackrel{4}{\hat{x}} \left. \!\!  \right. \!\!| \psi _0}\braket{\psi _0 |\!\! \stackrel{2}{\hat{v}} \left.\!\! \right. \!\! \stackrel{3}{\hat{w}} \left. \!\!  ^\dag  \right. \!\!| \psi _0} ,
\end{equation}
where we have put the numbers above each operator for each product, as obtained in figure \ref{fig:Pa_to_Pc}.
Since we know that, for all $p$, $q$, $r$, and $s$,
\begin{equation}
\left\langle \psi _0 \left|  \hat{p}^{\textcolor{white}{\dag}}\!\! \hat{q} \right| \psi _0\right\rangle = \left\langle \psi _0 \left| \hat{r}^\dag \hat{s}^\dag  \right| \psi _0\right\rangle = 0 , \quad
\left\langle \psi _0 \left| \hat{r} \hat{s}^\dag  \right| \psi _0\right\rangle = \delta _{rs}(1-n_r)(1-n_s) , \quad \left\langle \psi _0 \left| \hat{r}^\dag \hat{s}  \right| \psi _0\right\rangle = \delta _{rs} \, n_r n_s  ,
\end{equation}
where $n_s$ (respectively, $n_r$) is the occupation number (1 or 0) of $\varphi _s$ (respectively, $\varphi _r$) in $\ket{\psi _0}$, it follows that $\mathscr{I}_2$ can be simplified:
\begin{equation}
\mathscr{I}_2 = \underbrace{\braket{\psi _0 | \hat{u}^\dag\hat{v}|\psi_0}}_{\delta_{uv}n_u n_v}\underbrace{\braket{\psi _0 | \hat{w}^\dag \hat{x}| \psi _0}}_{\delta_{wx} n_w n_x} - \underbrace{\braket{\psi _0 | \hat{u}^\dag \hat{w}^\dag| \psi _0}}_{0}\underbrace{\braket{\psi _0 | \hat{v}\hat{x}|\psi _0}}_{0} + \underbrace{\braket{\psi _0 | \hat{u}^\dag\hat{x}|\psi_0}}_{\delta_{ux} n_u n_x}\underbrace{\braket{\psi _0 | \hat{v} \hat{w}^\dag| \psi _0}}_{\delta_{vw}(1-n_v) (1-n_w)},
\end{equation}
i.e.,
\begin{equation}
\mathscr{I}_2 = \left\langle \psi _0 \left| \hat{u}^\dag \hat{v} \hat{w}^\dag \hat{x} \right| \psi _0\right\rangle = \delta _{uv}\delta_{wx}n_u n_v n_w n_x + \delta_{ux}\delta_{vw}n_u n_x (1-n_v) (1-n_w). \label{eq:final_I2}
\end{equation}
In this example we have seen that from $(4-1)!! = 3$ products one reduces a four-operator integral such as \eqref{eq:I2} to $2! = 2$ products of Kronecker's deltas.

$\;$

\noindent \textbf{Six-operator integrals} 

$\;$

\begin{figure}[h!]
\begin{center}
\includegraphics[scale=1]{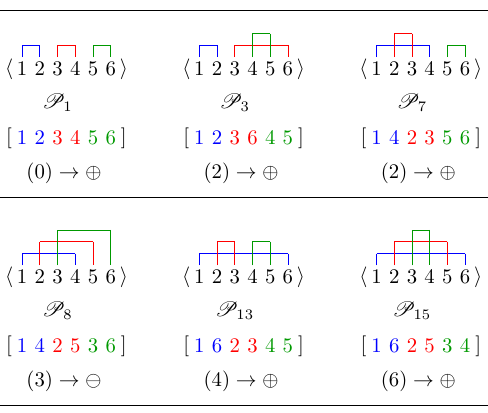}
\end{center}
\caption{Restriction of the decomposition of a six-operator integral to the sum of six integrals products $\mathscr{P}_{\ell}$. The number between parentheses is the number $f_\ell$ of times a pairing line crosses others (the number of times a line of a given color crosses lines with different colors) and the sign on the right of the arrow is the $\displaystyle (-1)^{f_\ell}$ signature. The reordered labels of the operators are reported between square brackets.}\label{fig:restricted_Pd_to_Pr}
\end{figure}

\noindent For six-operator integrals, we start again by applying the decomposition from Eq. \eqref{eq:initial_I}, implying $(6-1)!! = 15$ products of three two-operator integrals (See Appendix \ref{app:3-pair}) and, following the same considerations as for the four-operator integrals, we can reduce the sum of fifteen terms in Eq. \eqref{eq:initial_I} to the $3! = 6$ products of integrals in Eq. \eqref{eq:final_I}. These six products are reported in figure \ref{fig:restricted_Pd_to_Pr}.

\subsection{General equation-of-motion derivation of transition properties}

Let $\hat{\mathcal{T}}$ be a transition operator corresponding to the transition between two quantum states:
\begin{equation}
\hat{\mathcal{T}} = \ket{\psi _n}\bra{\psi _0} , \quad \hat{\mathcal{T}}^\dag =  \ket{\psi _0}\bra{\psi _n} ,
\end{equation}
with the constraint that the two states are normalized and orthogonal to each other (for example if they are eigenstates of the Hamiltonian operator)
\begin{equation}
\braket{\psi _0 | \psi _n} = \braket{\psi _n | \psi _0} = 0, \quad \braket{\psi _0 | \psi _0} = \braket{\psi _n | \psi _n} = 1 . \label{eq:eom-orth}
\end{equation}
We have that
\begin{equation}
\hat{\mathcal{T}} \ket{\psi _0} = \ket{\psi _n}\underbrace{\braket{\psi _0|\psi _0}}_{1} =  \ket{\psi _n}, \quad \bra{\psi _0}\hat{\mathcal{T}} ^\dag =\underbrace{\braket{\psi _0|\psi _0}}_{1}\bra{\psi _n} = \bra{\psi _n}, \label{eq:eom-proj}
\end{equation}
and
\begin{equation}
\hat{\mathcal{T}} \hat{\mathcal{T}} ^\dag = \ket{\psi _n}\bra{\psi _n} , \quad \hat{\mathcal{T}} ^\dag \hat{\mathcal{T}} = \ket{\psi _0}\bra{\psi _0} .
\end{equation}
It comes that
\begin{equation}
\left\langle \!\psi _0 \! \left| \left[ \hat{\mathcal{T}}^\dag , \left[ \hat{\mathcal{O}},\hat{\mathcal{T}}\right]\right] \right| \! \psi _0 \! \right\rangle = \underbrace{\braket{\psi _0 | \hat{\mathcal{T}}^\dag \hat{\mathcal{O}}\hat{\mathcal{T}} | \psi _0}}_{\displaystyle{\mathscr{D}^{\hat{\mathcal{O}}} _{\mathrm{I}}}} - \underbrace{\braket{\psi _0 | \hat{\mathcal{T}}^\dag \hat{\mathcal{T}} \hat{\mathcal{O}}| \psi _0}}_{\displaystyle{\mathscr{D}^{\hat{\mathcal{O}}} _{\mathrm{II}}}} - \underbrace{\braket{\psi _0 |  \hat{\mathcal{O}}\hat{\mathcal{T}} \hat{\mathcal{T}}^\dag | \psi _0}}_{\displaystyle{\mathscr{D}^{\hat{\mathcal{O}}} _{\mathrm{III}}}} + \underbrace{\braket{\psi _0 | \hat{\mathcal{T}}  \hat{\mathcal{O}}\hat{\mathcal{T}}^\dag| \psi _0}}_{\displaystyle{\mathscr{D}^{\hat{\mathcal{O}}} _{\mathrm{IV}}}} \label{eq:T_decomposition}
\end{equation}
can be simplified, according to Eqs. \eqref{eq:eom-orth} and \eqref{eq:eom-proj}, as
\begin{equation}
\left\langle \!\psi _0 \! \left| \left[ \hat{\mathcal{T}}^\dag , \left[ \hat{\mathcal{O}},\hat{\mathcal{T}}\right]\right] \right| \! \psi _0 \! \right\rangle =  \left\langle \!\psi _n \! \left| \hat{\mathcal{O}}\right| \! \psi _n \! \right\rangle - \left\langle \!\psi _0 \! \left| \hat{\mathcal{O}}\right| \! \psi _0 \! \right\rangle . \label{eq:eom-diff}
\end{equation}
Indeed, we have 
\begin{equation}
\mathscr{D}^{\hat{\mathcal{O}}} _{\mathrm{I}} = \braket{\psi _n | \hat{\mathcal{O}} | \psi _n}   , \quad \mathscr{D}^{\hat{\mathcal{O}}} _{\mathrm{II}} = \braket{\psi _0 | \hat{\mathcal{O}} | \psi _0}  ,
\end{equation}
and
\begin{equation}
\mathscr{D}^{\hat{\mathcal{O}}} _{\mathrm{III}} = \braket{\psi _0 | \hat{\mathcal{O}} | \psi _n}\braket{\psi _n | \psi _0} = 0 = \braket{\psi _0 | \psi _n}\braket{\psi _0 | \hat{\mathcal{O}} | \psi _0}\braket{\psi _n | \psi _0} = \mathscr{D}^{\hat{\mathcal{O}}} _{\mathrm{IV}} .
\end{equation}
If, instead of the difference of expectation value between two states for a given operator, we are interested in transition properties, those can be deduced by using
\begin{equation}
\left\langle \!\psi _0 \! \left|\left[ \hat{\mathcal{O}},\hat{\mathcal{T}}\right]  \right| \! \psi _0 \! \right\rangle = \underbrace{\left\langle \!\psi _0 \! \left| \hat{\mathcal{O}}\hat{\mathcal{T}}  \right| \! \psi _0 \! \right\rangle}_{\displaystyle{\mathscr{T}^{\hat{\mathcal{O}}}_\mathrm{I}}} - \underbrace{\left\langle \!\psi _0 \! \left| \hat{\mathcal{T}}\hat{\mathcal{O}}  \right| \! \psi _0 \! \right\rangle}_{\displaystyle{\mathscr{T}^{\hat{\mathcal{O}}}_{\mathrm{II}}}} = \left.\left. \left. \left\langle \!\psi _0 \! \left| \hat{\mathcal{O}}  \right| \! \psi _n \! \right\rangle \right\langle \! \psi _0  \right| \! \psi _0 \! \right\rangle -  \left\langle \! \psi _0  \left| \psi _n \! \left\rangle \left\langle \!\psi _0 \! \left| \hat{\mathcal{O}}  \right| \! \psi _0 \! \right\rangle \right. \right. \right.
\end{equation}
which reduces to
\begin{equation}
\left\langle \!\psi _0 \! \left|\left[ \hat{\mathcal{O}},\hat{\mathcal{T}}\right]  \right| \! \psi _0 \! \right\rangle = \left\langle \!\psi _0 \! \left| \hat{\mathcal{O}}  \right| \! \psi _n \! \right\rangle . \label{eq:transition_operators}
\end{equation}

\section{Evaluation of one-body density matrices using Wick's theorem}

\subsection{Hypotheses}

As we have seen before, the canonical space $\mathpzc{C}$ is divided into two sub-spaces: the occupied ($\mathpzc{O}$) and virtual ($\mathpzc{V}$) spaces
\begin{equation}\label{eq:ovspaces}
\mathpzc{O} = \left\lbrace \varphi _i  \; : \; 1 \leq i \leq N \right\rbrace , \; \mathpzc{V} = \left\lbrace \varphi _a  \; : \; N < a \leq L \right\rbrace .
\end{equation}
From now on, the second quantization operators with the letters $i$ or $j$ will point a spinorbital belonging to $\mathpzc{O}$ while letters $a$ and $b$ will point those from $\mathpzc{V}$, and $p$, $q$, $r$ and $s$ will be pointing any spinorbital of the canonical space.
In this section, the six-operator integrals (see Eq. \eqref{eq:I_M} with $M=3$) we will have to evaluate will involve two pairs of operators with each pair being the product of two operators belonging to two different subspaces ($\hat{a}^\dag \hat{i}$ or $\hat{j}^\dag \hat{b}$ for example) while the remaining pair of operators will be the $\hat{r}^\dag \hat{s}$ pair with $\varphi _r$ and $\varphi _s$ simply belonging to $\mathpzc{C}$ with no restriction. From this consideration we can write
\begin{equation}
\forall (\hat{q}, \hat{R}), \; \varphi_p  \in \mathpzc{V} \Longrightarrow \braket{ \psi _0 | \hat{p}^\dag \hat{q} \underbrace{\left({}^{[2]}\hat{c} _1 {}^{[2]}\hat{c}_{2}{}^{[3]}\hat{c} _1 {}^{[3]}\hat{c}_{2}\right)}_{\hat{R}} | \psi _0} = 0  \label{eq:th1a}
\end{equation}

and

\begin{equation}
\forall (\hat{p}^\dag, \hat{R}), \; \varphi_q  \in \mathpzc{V} \Longrightarrow  \braket{ \psi _0 | \underbrace{\left({}^{[1]}\hat{c} _1 {}^{[1]}\hat{c}_{2}{}^{[2]}\hat{c} _1 {}^{[2]}\hat{c}_{2}\right)}_{\hat{R}} \hat{p}^\dag \hat{q} | \psi _0} = 0. \label{eq:th1b}
\end{equation}
\noindent These rules can be deduced from a quick look at the structure of the products in figures \ref{fig:restricted_Pd_to_Pr} and \ref{fig:Pd_to_Pr} where we see that every product starts by ``$\langle 1$" and contains at least one integral ending by ``$6 \rangle$", or simply by considering the basic rules of second quantization recalled in equations \eqref{eq:SQzero1} and \eqref{eq:SQzero2}, and knowing that in the dual space a creation operator becomes an annihilation operator and \textit{vice versa}.

As a direct consequence, the substantial number of combinations of chains of second quantization operators we could meet can be easily sorted without evaluating the decomposition \eqref{eq:final_I} for each, so that only five of them will be useful in the following section. Those are identified in Table \ref{table:EV}, where the only non-vanishing products $\mathscr{P}_{\ell}$ from figure \ref{fig:restricted_Pd_to_Pr} are reported and highlighted by a ``$\checkmark$" symbol.

\begin{table}[h!]\label{table:EV}
\begin{center}
\begin{tabular}{p{3cm}|p{1cm}| p{1.5cm}|p{1.5cm}|p{1.5cm}|p{1.5cm}|p{1.5cm}}
EV & Label & $\mathscr{P}_3 \, (+)$ & $\mathscr{P}_{7} \, (+) $ & $\mathscr{P}_{8} \, (-) $ & $\mathscr{P}_{13} \, (+)$ & $\mathscr{P}_{15} \, (+)$ \\
\hline
$\braket{\mathrm{\psi _0 |\hat{i}^\dag\hat{a}\hat{\textbf{r}}^\dag\hat{\textbf{s}}\hat{b}^\dag\hat{j}}| \psi _0}$ & $\lambda$ & & &  \checkmark & \checkmark & \checkmark  \\
$\braket{\mathrm{\psi _0 |\hat{\textbf{r}}^\dag\hat{\textbf{s}}\hat{i}^\dag\hat{a}\hat{b}^\dag\hat{j}}| \psi _0}$ & $\delta  $  & \checkmark & &  &  \\
$\braket{\mathrm{\psi _0 |\hat{i}^\dag\hat{a}\hat{b}^\dag\hat{j}\hat{\textbf{r}}^\dag\hat{\textbf{s}}}| \psi _0}$ & $\beta $  & &\checkmark & & &   \\
\hdashline
$\braket{\mathrm{\psi _0|\hat{\textbf{r}}^\dag \hat{\textbf{s}}\hat{a}^\dag \hat{i}\hat{b}^\dag\hat{j}| \psi _0}}$ & $\xi$ & & & & & \\
$\braket{\mathrm{\psi _0|\hat{i}^\dag \hat{a}\hat{j}^\dag \hat{b}\hat{\textbf{r}}^\dag\hat{\textbf{s}}| \psi _0}}$ & $\chi$ & & & & & \\
\end{tabular}
\end{center}
\caption{The five typical six-operator integrals to evaluate in the determination of the one-body density matrices arising from single-excitation transition operators (coupled or not to deexcitation operators) in the single-reference equation-of-motion formalism we are using in this contribution.}
\end{table}
\noindent From Table \ref{table:EV}, we deduce that, relatively to $\lambda _{8}$,
\begin{equation}
   \braket{\psi _0 | \hat{i}^\dag\hat{s} | \psi _0}\braket{\psi _0 |\hat{a}\hat{b}^\dag | \psi _0}\braket{\psi _0 |\hat{{r}}^\dag \hat{j}| \psi _0} = \delta_{is} n_s \delta _{ab} \delta _{rj}n_r.
\end{equation}
Similarly, we have, for $\lambda _{13}$,
\begin{equation}
 \braket{\psi _0 | \hat{i}^\dag\hat{j} | \psi _0}\braket{\psi _0 |\hat{a}\hat{r}^\dag | \psi _0}\braket{\psi _0 |\hat{{s}} \hat{b}^\dag| \psi _0} = \delta _{ij}\delta _{ar}(1-n_r) \delta _{sb}(1-n_s)
\end{equation}
and for $\lambda _{15}$, we find
\begin{equation}
  \braket{\psi _0 | \hat{i}^\dag\hat{j} | \psi _0}\braket{\psi _0 |\hat{a}\hat{b}^\dag | \psi _0}\braket{\psi _0 |\hat{{r}}^\dag \hat{s}| \psi _0} = \delta _{ij}\delta _{ab} \delta _{rs}n_r.
\end{equation}
We also see that
\begin{equation}
\lambda _{15} \longleftrightarrow \delta _{3} \longleftrightarrow \beta _{7}, \label{eq:lambda_delta_zeta}
\end{equation}
so actually only the knowledge of the value of three integrals will be required for evaluating the matrix elements in this section.

\subsection{Coupled hole/particle and particle/hole (de)excitations}

For this first class of single-reference excited-state calculation methods, that we will denote here by an ``$\eta$" symbol, the transition operator is composed of a sum of single-excitation operators $\textbf{x}_{ia}\hat{a}^\dag\hat{i}$ added to the corresponding deexcitations $\textbf{y}_{ia}\hat{i}^\dag\hat{a}$
\begin{equation}
\hat{\mathcal{T}}_\eta = \sum _{i=1}^N \sum _{a=N+1}^L \left(\textbf{x}_{ia} \,\hat{a}^\dag\hat{i} - \textbf{y}_{ia} \,\hat{i}^\dag\hat{a}\right).
\end{equation}
Since the sum is running over all the possible spinorbitals of each sub-space, i.e., $i$ runs over all the occupied space and $a$ over all the virtual one, $\hat{\mathcal{T}}_\eta$ can be rewritten
\begin{equation}
\hat{\mathcal{T}}_\eta = \underbrace{\sum _{i=1}^N \sum _{a=N+1}^L \textbf{x}_{ia} \,\hat{a}^\dag\hat{i}}_{\displaystyle{\hat{\mathcal{T}}_x}}\;  - \; \underbrace{\sum _{j=1}^N \sum _{b=N+1}^L \textbf{y}_{jb} \,\hat{j}^\dag\hat{b}}_{\displaystyle{\hat{\mathcal{T}}_y}} . \label{eq:Teta}
\end{equation}
The \textbf{x} and \textbf{y} coefficients entering into the composition of $\hat{\mathcal{T}}_\eta$ are the components of a transition vector satisfying
\begin{equation}
\textbf{x}^\top_{ia} = \textbf{x}_{ia} , \quad \textbf{y}_{jb}^\top = \textbf{y}_{jb},
\end{equation}
so that
\begin{equation}
\hat{\mathcal{T}}_\eta ^\dag = \hat{\mathcal{T}}_x^\dag - \hat{\mathcal{T}}_y ^\dag = \sum _{i=1}^N \sum _{a=N+1}^L \textbf{x}_{ia}^\top \,\hat{i}^\dag\hat{a} -  \sum _{j=1}^N \sum _{b=N+1}^L \textbf{y}^\top_{jb} \,\hat{b}^\dag\hat{j}.
\end{equation}
While replacing $\hat{\mathcal{O}}$ in Eq. \eqref{eq:eom-diff} by the Hamiltonian leads naturally in the first quantization to the transition energy (see Appendix \ref{app:EOM}), the transition operator $\hat{\mathcal{T}}_\eta$ is sometimes inserted in Eq. \eqref{eq:eom-diff} as a substitution to $\hat{\mathcal{T}}$ for computing matrix elements. Therefore, in this contribution we will be interested in deriving the elements of the matrices (namely, the difference density matrix and transition density matrix) arising from the substitution of $\hat{\mathcal{T}}$ by $\hat{\mathcal{T}}_\eta$.

\subsubsection{Derivation of the difference density matrix elements}

The difference density matrix $\bm{\gamma}^\Delta$ is simply defined as the difference between two state density matrices, so according to Eq. \eqref{eq:dm_and_tdms}, if the transition operator $\hat{\mathcal{T}}$ was known exactly, replacing $\hat{\mathcal{O}}$ in Eq. \eqref{eq:eom-diff} by $\hat{r}^\dag \hat{s}$ would lead to the determination of the corresponding difference density matrix element
\begin{equation}
\left\langle \!\psi _0 \! \left| \left[ \hat{\mathcal{T}}^\dag  , \left[ \hat{r}^\dag \hat{s},\hat{\mathcal{T}}\right]\right] \right| \! \psi _0 \! \right\rangle = \left\langle \!\psi _n \! \left| \hat{r}^\dag \hat{s}\right| \! \psi _n \! \right\rangle - \left\langle \!\psi _0 \! \left| \hat{r}^\dag \hat{s}\right| \! \psi _0 \! \right\rangle = (\bm{\gamma}^\Delta)_{r,s} .
\end{equation}
 Here we will assess the approached value of these elements by inserting $\hat{r}^\dag\hat{s}$ and $\hat{\mathcal{T}}_\eta$ into Eq. \eqref{eq:eom-diff}:
\begin{equation}
(\bm{\gamma}^\Delta _\eta)_{r,s}  = \left\langle \!\psi _0 \! \left| \left[ \hat{\mathcal{T}}^\dag _\eta , \left[ \hat{r}^\dag \hat{s},\hat{\mathcal{T}}_\eta\right]\right] \right| \! \psi _0 \! \right\rangle .
\end{equation}
The $(\bm{\gamma}^\Delta _\eta)_{r,s}$ can be decomposed into four contributions, as in Eq. \eqref{eq:T_decomposition}:
\begin{equation}
(\bm{\gamma}^\Delta _\eta)_{r,s} = \mathscr{D}^{\hat{r}^\dag \! \hat{s}} _{\mathrm{I},\eta} -  \mathscr{D}^{\hat{r}^\dag \! \hat{s}} _{\mathrm{II},\eta} - \mathscr{D}^{\hat{r}^\dag \! \hat{s}} _{\mathrm{III},\eta} + \mathscr{D}^{\hat{r}^\dag \! \hat{s}} _{\mathrm{IV},\eta}
\end{equation}
leading to four matrix contributions to $\bm{\gamma}^\Delta$
\begin{equation}
\bm{\gamma}^\Delta _\eta = \mathscr{D}^{\Delta} _{\mathrm{I},\eta} -  \mathscr{D}^{\Delta} _{\mathrm{II},\eta} - \mathscr{D}^{\Delta} _{\mathrm{III},\eta} + \mathscr{D}^{\Delta} _{\mathrm{IV},\eta} \label{eq:gamma_Delta_structure}
\end{equation}
We will detail each of these four components by using the definition of $\hat{\mathcal{T}}_\eta$. For the first one, we have
\begin{equation}
\mathscr{D}^{\hat{r}^\dag \! \hat{s}} _{\mathrm{I},\eta} = \left\langle \!\psi _0 \! \left| \left( \hat{\mathcal{T}}_x ^\dag - \hat{\mathcal{T}}_y ^\dag \right) \hat{r}^\dag\hat{s}\left( \hat{\mathcal{T}}_x - \hat{\mathcal{T}}_y \right)  \right| \! \psi _0 \! \right\rangle = \mathscr{D}^{xx} _{\mathrm{I},rs} - \mathscr{D}^{yx} _{\mathrm{I},rs} - \mathscr{D}^{xy} _{\mathrm{I},rs}  + \mathscr{D}^{yy} _{\mathrm{I},rs},
\end{equation}
with

\begin{equation}
\mathscr{D}^{xx} _{\mathrm{I},rs} = \left\langle \!\psi _0 \! \left|   \hat{\mathcal{T}}_x ^\dag  \hat{r}^\dag\hat{s}  \hat{\mathcal{T}}_x   \right| \! \psi _0 \! \right\rangle = \sum _{i,j = 1}^N \sum _{a,b= N+1}^L \textbf{x}_{ia}^\top \textbf{x}_{jb} \braket{\psi _0 | \hat{i}^\dag \hat{a}\hat{r}^\dag\hat{s}\hat{b}^\dag \hat{j}| \psi _0}, \nonumber
\end{equation}

\begin{equation}
\mathscr{D}^{yx} _{\mathrm{I},rs} = \left\langle \!\psi _0 \! \left|   \hat{\mathcal{T}}_y ^\dag  \hat{r}^\dag\hat{s}  \hat{\mathcal{T}}_x   \right| \! \psi _0 \! \right\rangle = \sum _{i,j = 1}^N \sum _{a,b= N+1}^L \textbf{y}_{ia}^\top \textbf{x}_{jb} \braket{\psi _0 | \hat{a}^\dag \hat{i}\hat{r}^\dag\hat{s}\hat{b}^\dag \hat{j}| \psi _0}, \nonumber
\end{equation}

\begin{equation}
\mathscr{D}^{xy} _{\mathrm{I},rs} = \left\langle \!\psi _0 \! \left|   \hat{\mathcal{T}}_x ^\dag  \hat{r}^\dag\hat{s}  \hat{\mathcal{T}}_y   \right| \! \psi _0 \! \right\rangle = \sum _{i,j = 1}^N \sum _{a,b= N+1}^L \textbf{x}_{ia}^\top \textbf{y}_{jb} \braket{\psi _0 | \hat{i}^\dag \hat{a}\hat{r}^\dag\hat{s}\hat{j}^\dag \hat{b}| \psi _0}, \nonumber
\end{equation}

\begin{equation}
\mathscr{D}^{yy} _{\mathrm{I},rs} = \left\langle \!\psi _0 \! \left|   \hat{\mathcal{T}}_y ^\dag  \hat{r}^\dag\hat{s}  \hat{\mathcal{T}}_y   \right| \! \psi _0 \! \right\rangle = \sum _{i,j = 1}^N \sum _{a,b= N+1}^L \textbf{y}_{ia}^\top \textbf{y}_{jb} \braket{\psi _0 | \hat{a}^\dag \hat{i}\hat{r}^\dag\hat{s}\hat{j}^\dag \hat{b}| \psi _0}.
\end{equation}
According to equations \eqref{eq:th1a} and \eqref{eq:th1b}, we see that only the first component of $\mathscr{D}^{\hat{r}^\dag \! \hat{s}} _{\mathrm{I},\eta}$, is not vanishing. Indeed, the second component starts by $\hat{a}^\dag$, and the two last components end by $\hat{b}$. This consideration leads us to state that since we know that $\hat{\mathcal{T}}_x$ and $\hat{\mathcal{T}}_y^\dag$ both start with an $\hat{a}^\dag$ or $\hat{b}^\dag$ operator, with $\varphi _a  \in \mathpzc{V}$ and $\varphi _b  \in \mathpzc{V}$, and that $\hat{\mathcal{T}}_y$ and $\hat{\mathcal{T}}_x^\dag$ are ending with an $\hat{a}$ or a $\hat{b}$ operator, we have that
\begin{align}
0 &=\braket{\psi_0 | \hat{\mathcal{T}}_x \cdots |\psi _0} \nonumber \\&= \braket{\psi _0 | \hat{\mathcal{T}}_y^\dag \cdots | \psi _0} \nonumber \\ &= \braket{\psi _0 |\cdots \hat{\mathcal{T}}_y|\psi _0} \nonumber \\ &= \braket{\psi _0 |\cdots \hat{\mathcal{T}}_x^\dag|\psi _0}.\label{eq:corollaryT}
\end{align}

$\;$

\noindent This will be very helpful for considerably reducing the number of contributions to the difference density matrix to assess when evaluating $\mathscr{D}^{\hat{r}^\dag \! \hat{s}} _{\mathrm{II},\eta}$, $\mathscr{D}^{\hat{r}^\dag \! \hat{s}} _{\mathrm{III},\eta}$ and $\mathscr{D}^{\hat{r}^\dag \! \hat{s}} _{\mathrm{IV},\eta}$ later.

\noindent Before moving to the assessment of $\mathscr{D}^{\hat{r}^\dag \! \hat{s}} _{\mathrm{II},\eta}$, and in order to develop $\mathscr{D}^{xx} _{\mathrm{I},rs}$, we recast the \textbf{x} and \textbf{y} \textit{vectors} components into square \textit{matrices}:
\begin{equation}
\textbf{x}_{ia} = (\tilde{\textbf{X}})_{i,a} , \; \textbf{x}^\top _{ia} = (\tilde{\textbf{X}}\tmat)_{a,i} , \; \tilde{\textbf{X}} \in \mathbb{R}^{L\times L}.
\end{equation}
Similarly,
\begin{equation}
\textbf{y}_{ia} = (\tilde{\textbf{Y}})_{i,a} , \; \textbf{y}^\top _{ia} = (\tilde{\textbf{Y}}\tmat)_{a,i} , \; \tilde{\textbf{Y}} \in \mathbb{R}^{L\times L}.
\end{equation}
The $\tilde{\textbf{X}}$ and $\tilde{\textbf{Y}}$ matrices are in fact composed of four blocks (\textit{occupied} $\times$ \textit{occupied} ($o$), \textit{virtual} $\times$ \textit{virtual} ($v$), \textit{virtual} $\times$ \textit{occupied} ($vo$) and \textit{occupied} $\times$ \textit{virtual} ($ov$).) Since $\varphi _i  \in \mathpzc{O}$ and $\varphi _a  \in \mathpzc{V}$, the only non-zero elements from these matrices are in the $ov$ block. The $\tilde{\textbf{X}}$ and $\tilde{\textbf{Y}}$ matrices have the following structure:
 \begin{eqnarray}
\tilde{\textbf{X}} = \left( 
 \begin{array}{cc}
0_o  & \textbf{X}   \\
0_{vo}  &  0_v \\
 \end{array}\right) , \; \tilde{\textbf{X}}\tmat = \left( 
 \begin{array}{cc}
0_o  & 0_{ov}   \\
\textbf{X}\tmat  &  0_v \\
 \end{array}\right) , \; \tilde{\textbf{X}}\tilde{\textbf{X}}\tmat = \textbf{XX}\tmat \oplus 0_v \; , \; \tilde{\textbf{X}}\tmat \tilde{\textbf{X}} = 0_o \oplus \textbf{X}\tmat \textbf{X} \nonumber
\end{eqnarray}
with
\begin{equation}
\textbf{X} \textbf{X}\tmat \in \mathbb{R}^{N\times N}, \; \textbf{X}\tmat \textbf{X} \in \mathbb{R}^{(L-N)\times(L-N)}
\end{equation}
and the zero matrices
$0_o \,(N\times N)$, $0_{vo} \,((L-N) \times N)$, $0_{ov} \,(N \times (L-N))$, and $0_v \,((L-N) \times (L-N))$.

We also have
 \begin{eqnarray}
\tilde{\textbf{Y}} = \left( 
 \begin{array}{cc}
0_o  & \textbf{Y}   \\
0_{vo}  &  0_v \\
 \end{array}\right) , \; \tilde{\textbf{Y}}\tmat = \left( 
 \begin{array}{cc}
0_o  & 0_{ov}   \\
\textbf{Y}\tmat  &  0_v \\
 \end{array}\right) , \; \tilde{\textbf{Y}}\tilde{\textbf{Y}}\tmat = \textbf{YY}\tmat \oplus 0_v \; , \; \tilde{\textbf{Y}}\tmat \tilde{\textbf{Y}} = 0_o \oplus \textbf{Y}\tmat \textbf{Y} \nonumber
\end{eqnarray}
with
\begin{equation}
\textbf{Y} \textbf{Y}\tmat \in \mathbb{R}^{N\times N} , \; \textbf{Y}\tmat \textbf{Y} \in \mathbb{R}^{(L-N)\times(L-N)}.
\end{equation}
If we take the $\mathscr{D}^{xx} _{\mathrm{I},rs}$ integrals, we see that they have the structure of the $\lambda$ integral in Table \ref{table:EV}, which means that $\mathscr{D}^{xx} _{\mathrm{I},rs}$ can be rewritten
\begin{align}
\mathscr{D}^{xx} _{\mathrm{I},rs} &= \sum _{i,j = 1}^N \sum _{a,b= N+1}^L \textbf{x}_{ia}^\top \textbf{x}_{jb} \underbrace{\braket{\psi _0 | \hat{i}^\dag \hat{a}\hat{r}^\dag\hat{s}\hat{b}^\dag \hat{j}| \psi _0}}_{\displaystyle{\lambda}} \nonumber \\
&= \sum _{i,j = 1}^N \sum _{a,b= N+1}^L (\tilde{\textbf{X}}\tmat)_{a,i} (\tilde{\textbf{X}})_{j,b} \left( - \delta _{is}n_s \delta_{ab}\delta_{rj}n_r + \delta_{ij}\delta_{ar}(1-n_r)\delta_{sb}(1-n_s) + \delta_{ij}\delta_{ab}\delta_{rs}n_r \right). \label{eq:Dxx_decomposition}
\end{align}
If we consider the first term, we have
\begin{equation}
- \sum _{i,j = 1}^N \sum _{a,b= N+1}^L (\tilde{\textbf{X}}\tmat)_{a,i} (\tilde{\textbf{X}})_{j,b} \, \delta _{is}n_s \delta_{ab}\delta_{rj}n_r = - \sum _{a,b=N+1}^L(\tilde{\textbf{X}}\tmat)_{a,s} (\tilde{\textbf{X}})_{r,b} \, \delta _{ab} n_r n_s = - \sum _{a=N+1}^L(\tilde{\textbf{X}}\tmat)_{a,s} (\tilde{\textbf{X}})_{r,a}\, n_r n_s.
\end{equation}
Since any $(\tilde{\textbf{X}})_{r,q}$ and $(\tilde{\textbf{X}}\tmat)_{q,s}$ matrix element with $q \leq N$ is zero, we can expand the sum over $a$ to any value between $1$ and $L$ without affecting the result
\begin{equation}
- \sum _{a=N+1}^L(\tilde{\textbf{X}})_{r,a}(\tilde{\textbf{X}}\tmat)_{a,s} \, n_r n_s= - \sum _{q=1}^L(\tilde{\textbf{X}})_{r,q}(\tilde{\textbf{X}}\tmat)_{q,s} n_r n_s = - (\tilde{\textbf{X}}\tilde{\textbf{X}}\tmat)_{r,s} \, n_r n_s . \label{eq:Dxx_term1}
\end{equation}
At this stage, we can drop the $n_r$ and $n_s$ factors since for any $r$ or $s$ superior to $N$ the $(\tilde{\textbf{X}}\tilde{\textbf{X}}\tmat)_{r,s}$ matrix element is zero.

For the second term of Eq. \eqref{eq:Dxx_decomposition}, we have
\begin{equation}
\sum _{i,j = 1}^N \sum _{a,b= N+1}^L (\tilde{\textbf{X}}\tmat)_{a,i} (\tilde{\textbf{X}})_{j,b} \, \delta_{ij}\delta_{ar}(1-n_r)\delta_{sb}(1-n_s) = \sum _{i = 1}^N  (\tilde{\textbf{X}}\tmat)_{r,i} (\tilde{\textbf{X}})_{i,s} \, (1-n_r)(1-n_s).
\end{equation}
For the same reason as for the first term, we see that any $(\tilde{\textbf{X}})_{p,s}$ and $(\tilde{\textbf{X}}\tmat)_{r,p}$ matrix element with $p > N$ is zero, so we can expand the sum over $i$ to any value between $1$ and $L$ without affecting the result
\begin{equation}
\sum _{p = 1}^L  (\tilde{\textbf{X}}\tmat)_{r,p} (\tilde{\textbf{X}})_{p,s} \, (1-n_r)(1-n_s) = (\tilde{\textbf{X}}\tmat \tilde{\textbf{X}})_{r,s}  \, (1-n_r)(1-n_s) \label{eq:Dxx_term2}
\end{equation}
and here the $(1-n_r)(1-n_s)$ factor can also be dropped since the $(\tilde{\textbf{X}}\tmat \tilde{\textbf{X}})_{r,s}$ matrix elements are vanishing for any $r$ or $s$ inferior or equal to $N$.

Finally, the third term in Eq. \eqref{eq:Dxx_decomposition} can be simplified into
\begin{equation}
\sum _{i,j = 1}^N \sum _{a,b= N+1}^L (\tilde{\textbf{X}}\tmat)_{a,i} (\tilde{\textbf{X}})_{j,b} \,\delta_{ij}\delta_{ab}\delta_{rs}n_r  = \sum _{i = 1}^N \sum _{a= N+1}^L (\tilde{\textbf{X}}\tmat)_{a,i} (\tilde{\textbf{X}})_{i,a} \,\delta_{rs}n_r =  \underbrace{\mathrm{tr}(\tilde{\textbf{X}}\tmat \tilde{\textbf{X}})}_{\displaystyle{\vartheta _x}}\delta_{rs}n_r . \label{eq:Dxx_term3}
\end{equation}
Here the $n_r$ factor cannot be dropped, as it is preceded by Kronecker's delta multiplied by a scalar factor, so the $n_r$ factor is here to ensure that this scalar factor will be multiplying the identity matrix $I_o$ covering the \textit{occupied} $\times$ \textit{occupied} ($o$) block in the difference density matrix.

We can therefore conclude from Eqs. \eqref{eq:Dxx_term1}, \eqref{eq:Dxx_term2} and \eqref{eq:Dxx_term3} that the first contribution to the difference density matrix reads
\begin{equation}
\mathscr{D}^{\hat{r}^\dag \! \hat{s}} _{\mathrm{I},\eta} = \vartheta _x\delta_{rs}n_r - (\tilde{\textbf{X}}\tilde{\textbf{X}}\tmat)_{r,s} + (\tilde{\textbf{X}}\tmat \tilde{\textbf{X}})_{r,s} \label{eq:DI_eta}
\end{equation}
which gives the following matrix contribution to $\bm{\gamma}^\Delta$:
\begin{equation}
\mathscr{D}^{\Delta} _{\mathrm{I},\eta} = \left(\vartheta _x I_o - {\textbf{X}}{\textbf{X}}\tmat\right) \oplus {\textbf{X}}\tmat{\textbf{X}}
\end{equation}
with the identity matrix $I_o \, (N\times N)$ spanning the occupied space. We now turn to the evaluation of $\mathscr{D}^{\hat{r}^\dag \! \hat{s}} _{\mathrm{II},\eta}$ which, according to Eq. \eqref{eq:T_decomposition} in which we shall use again $\hat{r}^\dag \hat{s}$ as $\hat{\mathcal{O}}$ and substitute again $\hat{\mathcal{T}}$ by $\hat{\mathcal{T}}_\eta$, reads
\begin{equation}
\mathscr{D}^{\hat{r}^\dag \! \hat{s}} _{\mathrm{II},\eta} = \left\langle \!\psi _0 \! \left|\left( \hat{\mathcal{T}}_x ^\dag - \hat{\mathcal{T}}_y^\dag\right)\left( \hat{\mathcal{T}}_x - \hat{\mathcal{T}}_y\right) \hat{r}^\dag\hat{s}  \right| \! \psi _0 \! \right\rangle .
\end{equation}
This contribution, according to equation \eqref{eq:corollaryT}, reduces to
\begin{equation}
\mathscr{D}^{\hat{r}^\dag \! \hat{s}} _{\mathrm{II},\eta} = \left\langle \!\psi _0 \! \left|\left( \hat{\mathcal{T}}_x ^\dag\right)\left( \hat{\mathcal{T}}_x - \hat{\mathcal{T}}_y\right) \hat{r}^\dag\hat{s}  \right| \! \psi _0 \! \right\rangle
\end{equation}
so we only have to assess two six-operator integrals:
\begin{equation}
\mathscr{D}^{\hat{r}^\dag \! \hat{s}} _{\mathrm{II},\eta} = \underbrace{\left\langle \!\psi _0 \! \left| \hat{\mathcal{T}}_x ^\dag \hat{\mathcal{T}}_x  \hat{r}^\dag\hat{s}  \right| \! \psi _0 \! \right\rangle}_{\displaystyle{\mathscr{D}^{xx} _{\mathrm{II},rs}}} - \underbrace{\left\langle \!\psi _0 \! \left| \hat{\mathcal{T}}_x ^\dag \hat{\mathcal{T}}_y  \hat{r}^\dag\hat{s}  \right| \! \psi _0 \! \right\rangle}_{\displaystyle{\mathscr{D}^{xy} _{\mathrm{II},rs}}},
\end{equation}
with, according to Table \ref{table:EV},
\begin{equation}
\mathscr{D}^{xx} _{\mathrm{II},rs} = \sum _{i,j = 1}^N \sum _{a,b= N+1}^L \textbf{x}_{ia}^\top \textbf{x}_{jb} \underbrace{\braket{\psi _0 | \hat{i}^\dag \hat{a}\hat{b}^\dag \hat{j}\hat{r}^\dag\hat{s}| \psi _0}}_{\displaystyle{\beta}} = \vartheta _x \delta _{rs} n_r
\end{equation}
and
\begin{equation}
\mathscr{D}^{xy} _{\mathrm{II},rs} = \sum _{i,j = 1}^N \sum _{a,b= N+1}^L \textbf{x}_{ia}^\top \textbf{y}_{jb} \underbrace{\braket{\psi _0 | \hat{i}^\dag \hat{a}\hat{j}^\dag \hat{b}\hat{r}^\dag\hat{s}| \psi _0}}_{\displaystyle{\chi}} = 0
\end{equation}
so the second contribution to the difference density matrix is, according to Eqs. \eqref{eq:lambda_delta_zeta} and \eqref{eq:Dxx_term3},
\begin{equation}
\mathscr{D}^{\Delta} _{\mathrm{II},\eta} = \vartheta _x I_o \oplus 0_v .
\end{equation}
The matrix elements of the third contribution to $\bm{\gamma}^\Delta$, $\mathscr{D}^{\hat{r}^\dag \! \hat{s}} _{\mathrm{III},\eta}$, reads
\begin{equation}
\mathscr{D}^{\hat{r}^\dag \! \hat{s}} _{\mathrm{III},\eta} = \left\langle \!\psi _0 \! \left| \hat{r}^\dag\hat{s} \left( \hat{\mathcal{T}}_x - \hat{\mathcal{T}}_y \right)\left( \hat{\mathcal{T}}_x^\dag - \hat{\mathcal{T}}^\dag _y\right) \right| \! \psi _0 \! \right\rangle .
\end{equation}
From equation \eqref{eq:corollaryT}, it reduces to
\begin{equation}
\mathscr{D}^{\hat{r}^\dag \! \hat{s}} _{\mathrm{III},\eta} = \underbrace{\left\langle \!\psi _0 \! \left| \hat{r}^\dag\hat{s}\hat{\mathcal{T}}_y  \hat{\mathcal{T}}_y^\dag    \right| \! \psi _0 \! \right\rangle}_{\displaystyle{\mathscr{D}^{yy} _{\mathrm{III},rs}}} - \underbrace{\left\langle \!\psi _0 \! \left| \hat{r}^\dag\hat{s} \hat{\mathcal{T}}_x  \hat{\mathcal{T}}_y ^\dag   \right| \! \psi _0 \! \right\rangle}_{\displaystyle{\mathscr{D}^{xy} _{\mathrm{III},rs}}},
\end{equation}
with, according to Table \ref{table:EV},
\begin{equation}
\mathscr{D}^{yy} _{\mathrm{III},rs} = \sum _{i,j = 1}^N \sum _{a,b= N+1}^L \textbf{y}_{ia} \textbf{y}_{jb}^\top \underbrace{\braket{\psi _0 | \hat{r}^\dag \hat{s}\hat{i}^\dag \hat{a}\hat{b}^\dag\hat{j}| \psi _0}}_{\displaystyle{\delta}} = \underbrace{\mathrm{tr}\left( \tilde{\textbf{Y}}\tilde{\textbf{Y}}\tmat\right)}_{\displaystyle{\vartheta_y}} \delta_{rs}n_r
\end{equation}
and
\begin{equation}
\mathscr{D}^{xy} _{\mathrm{III},rs} = \sum _{i,j = 1}^N \sum _{a,b= N+1}^L \textbf{x}_{ia} \textbf{y}_{jb}^\top \underbrace{\braket{\psi _0 | \hat{r}^\dag \hat{s}\hat{a}^\dag \hat{i}\hat{b}^\dag\hat{j}| \psi _0}}_{\displaystyle{\xi}} = 0
\end{equation}
so the third contribution to the difference density matrix is, according to Eqs. \eqref{eq:lambda_delta_zeta} and \eqref{eq:Dxx_term3},
\begin{equation}
\mathscr{D}^{\Delta} _{\mathrm{III},\eta} = \vartheta _y I_0 \oplus 0_v .
\end{equation}
Finally, the matrix elements of the fourth contribution to $\bm{\gamma}^\Delta$, $\mathscr{D}^{\hat{r}^\dag \! \hat{s}} _{\mathrm{IV},\eta}$, read
\begin{equation}
\mathscr{D}^{\hat{r}^\dag \! \hat{s}} _{\mathrm{IV},\eta} = \left\langle \!\psi _0 \! \left|  \left( \hat{\mathcal{T}}_x - \hat{\mathcal{T}}_y \right)\hat{r}^\dag\hat{s}\left( \hat{\mathcal{T}}_x^\dag - \hat{\mathcal{T}}^\dag _y\right) \right| \! \psi _0 \! \right\rangle .
\end{equation}
From equation \eqref{eq:corollaryT}, it reduces to
\begin{equation}
\mathscr{D}^{\hat{r}^\dag \! \hat{s}} _{\mathrm{IV},\eta} = {\left\langle \!\psi _0 \! \left| \hat{\mathcal{T}}_y  \hat{r}^\dag\hat{s} \hat{\mathcal{T}}_y^\dag    \right| \! \psi _0 \! \right\rangle} = \mathscr{D}^{yy} _{\mathrm{IV},rs}
\end{equation}
with, according to Table \ref{table:EV},
\begin{equation}
\mathscr{D}^{yy} _{\mathrm{IV},rs} = \sum _{i,j = 1}^N \sum _{a,b= N+1}^L \textbf{y}_{ia} \textbf{y}_{jb}^\top \underbrace{\braket{\psi _0 |\hat{i}^\dag \hat{a} \hat{r}^\dag \hat{s}\hat{b}^\dag\hat{j}| \psi _0}}_{\displaystyle{\lambda}} = \vartheta _y\delta_{rs}n_r - (\tilde{\textbf{Y}}\tilde{\textbf{Y}}\tmat)_{r,s} + (\tilde{\textbf{Y}}\tmat \tilde{\textbf{Y}})_{r,s}  
\end{equation}
which leads to the following matrix contribution to $\bm{\gamma}^\Delta$:
\begin{equation}
\mathscr{D}^{\Delta} _{\mathrm{IV},\eta} = \left(\vartheta _y I_o - {\textbf{Y}}{\textbf{Y}}\tmat \right) \oplus {\textbf{Y}}\tmat{\textbf{Y}},
\end{equation} 
the last equalities being deduced by simply transferring the derivation of Eq. \eqref{eq:DI_eta} to $\mathscr{D}^{yy} _{\mathrm{IV},rs}$. 

We finally conclude from Eq. \eqref{eq:gamma_Delta_structure}, by combining the different results we just derived, that the difference density matrix $\bm{\gamma}^\Delta$ has the following structure:
\begin{equation}
\bm{\gamma}^\Delta _\eta = \underbrace{\left(\vartheta _x I_o - {\textbf{X}}{\textbf{X}}\tmat\right) \oplus {\textbf{X}}\tmat{\textbf{X}}}_{\displaystyle{\mathscr{D}^{\Delta} _{\mathrm{I},\eta}}} - \overbrace{\left(\vartheta _x I_o \oplus 0_v\right)}^{\displaystyle{\mathscr{D}^{\Delta} _{\mathrm{II},\eta}}} - \underbrace{\left(\vartheta _y I_o \oplus 0_v\right)}_{\displaystyle{\mathscr{D}^{\Delta} _{\mathrm{III},\eta}}}  + \overbrace{\left(\vartheta _y I_o - {\textbf{Y}}{\textbf{Y}}\tmat\right) \oplus {\textbf{Y}}\tmat{\textbf{Y}}}^{\displaystyle{\mathscr{D}^{\Delta} _{\mathrm{IV},\eta}}},
\end{equation}
i.e.,
\begin{equation}
  \left(- \textbf{X}{\textbf{X}}\tmat - \textbf{Y}\textbf{Y}\tmat \right) \oplus \left({\textbf{X}}\tmat{\textbf{X}} + {\textbf{Y}}\tmat{\textbf{Y}} \right) = \bm{\gamma}^\Delta_\eta \in \mathbb{R}^{L \times L}. \label{eq:DDM-eta}
\end{equation}

\subsubsection{Derivation of the transition density matrix elements}

According to Eqs. \eqref{eq:dm_and_tdms} and \eqref{eq:transition_operators}, the transition density matrix elements write
\begin{equation}
(\bm{\gamma}^{0 \rightarrow n} _\eta)_{s,r} = \left\langle \!\psi _0 \! \left|\left[ \hat{r}^\dag\hat{s},\hat{\mathcal{T}}_\eta\right]  \right| \! \psi _0 \! \right\rangle = \underbrace{\left\langle \!\psi _0 \! \left| \hat{r}^\dag\hat{s}\hat{\mathcal{T}}_\eta  \right| \! \psi _0 \! \right\rangle}_{\displaystyle{\mathscr{T}^{\hat{r}^\dag \! \hat{s}}_{\mathrm{I},\eta}}} - \underbrace{\left\langle \!\psi _0 \! \left| \hat{\mathcal{T}}_\eta\hat{r}^\dag \hat{s}  \right| \! \psi _0 \! \right\rangle}_{\displaystyle{\mathscr{T}^{\hat{r}^\dag \!\hat{s}}_{\mathrm{II},\eta}}}
\end{equation}
with
\begin{equation}
\mathscr{T}^{\hat{r}^\dag \! \hat{s}}_{\mathrm{I},\eta} = \left\langle \!\psi _0 \! \left| \hat{r}^\dag\hat{s}\left( \hat{\mathcal{T}}_x - \hat{\mathcal{T}}_y\right)  \right| \! \psi _0 \! \right\rangle = \left\langle \!\psi _0 \! \left| \hat{r}^\dag\hat{s} \hat{\mathcal{T}}_x   \right| \! \psi _0 \! \right\rangle
\end{equation}
that we can rewrite
\begin{equation}
\mathscr{T}^{\hat{r}^\dag \! \hat{s}}_{\mathrm{I},\eta} = \sum _{i=1}^N \sum _{a=N+1}^L \textbf{x}_{ia} \left\langle \!\psi _0 \! \left| \hat{r}^\dag\hat{s}\hat{a}^\dag \hat{i} \right| \! \psi _0 \! \right\rangle = \sum _{i=1}^N \sum _{a=N+1}^L \textbf{x}_{ia} \left( \delta _{rs}\delta_{ai} + \delta _{ri}\delta _{sa} \right) = \sum _{i=1}^N \sum _{a=N+1}^L \textbf{x}_{ia} \delta _{ri}n_r \delta _{sa}(1-n_s).
\end{equation}
Similarly, for $\mathscr{T}^{\hat{r}^\dag \! \hat{s}}_{\mathrm{II},\eta}$, we find
\begin{equation}
\mathscr{T}^{\hat{r}^\dag \! \hat{s}}_{\mathrm{II},\eta} = \left\langle \!\psi _0 \! \left| \left( \hat{\mathcal{T}}_x - \hat{\mathcal{T}}_y\right) \hat{r}^\dag\hat{s}  \right| \! \psi _0 \! \right\rangle =-  \left\langle \!\psi _0 \! \left| \hat{\mathcal{T}}_y \hat{r}^\dag\hat{s}  \right| \! \psi _0 \! \right\rangle ,
\end{equation}
i.e., 
\begin{equation}
\mathscr{T}^{\hat{r}^\dag \! \hat{s}}_{\mathrm{II},\eta} = - \sum _{i=1}^N \sum _{a=N+1}^L \textbf{y}_{ia} \left\langle \!\psi _0 \! \left| \hat{i}^\dag \hat{a}\hat{r}^\dag\hat{s} \right| \! \psi _0 \! \right\rangle = - \sum _{i=1}^N \sum _{a=N+1}^L \textbf{y}_{ia} \left( \delta _{ia}\delta_{rs} + \delta _{is}\delta _{ar} \right) = - \sum _{i=1}^N \sum _{a=N+1}^L \textbf{y}_{ia} \delta _{is}n_s \delta _{ar}(1-n_r).
\end{equation}
The combination of $\mathscr{T}^{\hat{r}^\dag \! \hat{s}}_{\mathrm{I},\eta}$ and $\mathscr{T}^{\hat{r}^\dag \! \hat{s}}_{\mathrm{II},\eta}$ leads finally to the transition density matrix
\begin{eqnarray}
  \left( 
 \begin{array}{cc}
0_o  & \textbf{Y}   \\
\textbf{X}\tmat  &  0_v \\
 \end{array}\right) = \bm{\gamma} ^{0\rightarrow n}_\eta \in \mathbb{R}^{L \times L} \label{eq:TDM-eta}
\end{eqnarray}
which is not Hermitian.
\subsection{Comment on the nature of the transition operators}

This section gives a comment related to the consequences of substituting $\hat{\mathcal{T}}$ by $\hat{\mathcal{T}}_\eta$ in the Eq. \eqref{eq:eom-diff} and \eqref{eq:transition_operators}. While this substitution is sometimes introduced to approach the transition energy and to derive properties and objects related to the transition, one should be careful about the interpretation given to this substitution. Indeed, in the $\hat{\mathcal{T}}_\eta$ operator, one could interpret the deexcitation part $\hat{\mathcal{T}}_y$ as a correction to the full CI truncation to single excitations that constitutes $\hat{\mathcal{T}}_x$ for deriving the properties and objects related above. Due to the substitution of $\hat{\mathcal{T}}$ by $\hat{\mathcal{T}}_\eta$ for accomplishing such operations, one could also be tempted to interpret $\hat{\mathcal{T}}_\eta$ as being an approximation to $\hat{\mathcal{T}}$. Indeed, if we consider, in the first quantization, that, in the EOM theory,
\begin{equation}
\braket{\psi _0 | \psi _0} = 1 \Longleftrightarrow \left(\left[\left\langle \!\psi _0 \! \left|\hat{\mathcal{T}}  \right| \! \psi _0 \! \right\rangle = \braket{\psi _0 | \psi _n}\braket{\psi _0 | \psi _0} = 0\right] \, \Longleftrightarrow \, \braket{\psi _0 | \psi _n} = 0 \right), \label{eq:orthogonality}
\end{equation}
and notice that, according to equation \eqref{eq:corollaryT},
\begin{equation}
\left\langle \!\psi _0 \! \left|\hat{\mathcal{T}}_\eta  \right| \! \psi _0 \! \right\rangle = \left\langle \!\psi _0 \! \left|\hat{\mathcal{T}}_x - \hat{\mathcal{T}}_y   \right| \! \psi _0 \! \right\rangle = \underbrace{\left\langle \!\psi _0 \! \left|\hat{\mathcal{T}}_x    \right| \! \psi _0 \! \right\rangle}_{0} - \underbrace{\left\langle \!\psi _0 \! \left|  \hat{\mathcal{T}}_y   \right| \! \psi _0 \! \right\rangle}_{0} = 0; \label{eq:orthogonality_eta}
\end{equation}
if we also consider that, given that the two states are orthogonal, in the EOM theory, we find
\begin{equation}
\left\langle \!\psi _0 \! \left|\left[ \hat{\mathcal{T}}^\dag,\hat{\mathcal{T}}\right]  \right| \! \psi _0 \! \right\rangle = \braket{\psi _0 | \psi _0} \braket{\psi _n | \psi_n}\braket{\psi _0 | \psi _0} - \braket{\psi _0 | \psi _n}\braket{\psi _0 | \psi _0}\braket{\psi _n |\psi _0} = \braket{\psi _n | \psi _n} - 0 = 1,
\end{equation}
i.e.,
\begin{equation}
 \left\langle \!\psi _0 \! \left|\left[ \hat{\mathcal{T}}^\dag,\hat{\mathcal{T}}\right]  \right| \! \psi _0 \! \right\rangle = 1 \; \Longleftrightarrow \; \braket{\psi _n | \psi _n} = 1,  \label{eq:normalization}
\end{equation}
and notice that
\begin{equation}
 \left\langle \!\psi _0 \! \left|\left[ \hat{\mathcal{T}}_\eta ^\dag,\hat{\mathcal{T}}_\eta \right]  \right| \! \psi _0 \! \right\rangle =  \left\langle \!\psi _0 \! \left|\left( \hat{\mathcal{T}}_x^\dag - \hat{\mathcal{T}}_y^\dag \right) \left( \hat{\mathcal{T}}_x - \hat{\mathcal{T}}_y \right)  \right| \! \psi _0 \! \right\rangle - \left\langle \!\psi _0 \! \left|\left( \hat{\mathcal{T}}_x - \hat{\mathcal{T}}_y \right)\left( \hat{\mathcal{T}}_x^\dag - \hat{\mathcal{T}}_y^\dag \right)  \right| \! \psi _0 \! \right\rangle 
\end{equation}
which reduces, due to equation \eqref{eq:corollaryT}, to
\begin{equation}
 \left\langle \!\psi _0 \! \left|\left[ \hat{\mathcal{T}}_\eta ^\dag,\hat{\mathcal{T}}_\eta \right]  \right| \! \psi _0 \! \right\rangle =  \left\langle \!\psi _0 \! \left| \hat{\mathcal{T}}_x^\dag  \hat{\mathcal{T}}_x  \right| \! \psi _0 \! \right\rangle - \left\langle \!\psi _0 \! \left| \hat{\mathcal{T}}_y   \hat{\mathcal{T}}_y^\dag    \right| \! \psi _0 \! \right\rangle ,
\end{equation}
with
\begin{equation}
\left\langle \!\psi _0 \! \left| \hat{\mathcal{T}}_x^\dag  \hat{\mathcal{T}}_x  \right| \! \psi _0 \! \right\rangle =  \sum _{i,j = 1}^N \sum _{a,b= N+1}^L \textbf{x}_{ia}^\top \textbf{x}_{jb} \braket{\psi _0 | \hat{i}^\dag \hat{a}\hat{b}^\dag \hat{j}| \psi _0} = \sum _{i,j = 1}^N \sum _{a,b= N+1}^L (\tilde{\textbf{X}}\tmat)_{a,i}(\tilde{\textbf{X}})_{j,b} \left( \delta_{ia}\delta _{bj} + \delta_{ij}\delta_{ab} \right)  = \vartheta _x
\end{equation}
and
\begin{equation}
\left\langle \!\psi _0 \! \left| \hat{\mathcal{T}}_y  \hat{\mathcal{T}}_y^\dag  \right| \! \psi _0 \! \right\rangle =  \sum _{i,j = 1}^N \sum _{a,b= N+1}^L \textbf{y}_{ia} \textbf{y}_{jb}^\top \braket{\psi _0 | \hat{i}^\dag \hat{a}\hat{b}^\dag \hat{j}| \psi _0} = \sum _{i,j = 1}^N \sum _{a,b= N+1}^L (\tilde{\textbf{Y}})_{i,a}(\tilde{\textbf{Y}}\tmat)_{b,j} \left( \delta_{ia}\delta _{bj} + \delta_{ij}\delta_{ab} \right)  = \vartheta _y,
\end{equation}
so
\begin{equation}
\left\langle \!\psi _0 \! \left|\left[ \hat{\mathcal{T}}_\eta ^\dag,\hat{\mathcal{T}}_\eta \right]  \right| \! \psi _0 \! \right\rangle = \vartheta _x - \vartheta _y \equiv \textbf{x}\tvec  \textbf{x} - \textbf{y}\tvec  \textbf{y} = 1 \label{eq:normalization_eta}
\end{equation}
where $\textbf{x}\tvec \textbf{x} - \textbf{y}\tvec \textbf{y}$ is known to be equal to unity, due to the non-standard normalization conditions of the \textbf{x} and \textbf{y} couple of vectors, we find quite good similarities between the first quantization EOM $\hat{\mathcal{T}}$ properties and the properties of $\hat{\mathcal{T}}_\eta$.

Indeed, putting in parallel results in Eq. \eqref{eq:orthogonality} with that of Eq. \eqref{eq:orthogonality_eta}, and the one in Eq. \eqref{eq:normalization} with the one from \eqref{eq:normalization_eta}, one could be tempted to conclude that the quantum state $\ket{\psi _{n,\eta}}$ we could obtain by operating $\hat{\mathcal{T}}_\eta$ on $\ket{\psi_0}$ is orthogonal to $\ket{\psi_0}$ and is normalized. However, one immediately sees that when acting on $\ket{\psi_0}$, $\hat{\mathcal{T}}_\eta$ is not nilpotent, unlike $\hat{\mathcal{T}}$: while
\begin{equation}
\hat{\mathcal{T}}\hat{\mathcal{T}} = \ket{\psi _n}\braket{\psi _0 |\psi _n} \bra{\psi _0}
\end{equation}
is always zero, the ground state expectation value of $\hat{\mathcal{T}}_\eta \hat{\mathcal{T}}_\eta$ reduces to the ground state expectation value of 
\begin{equation}
 - \hat{\mathcal{T}}_y\hat{\mathcal{T}}_x =  - \sum _{i,j=1}^N \sum _{a,b=N+1}^L \textbf{y}_{ia}\textbf{x}_{jb} \hat{i}^\dag\hat{a}\hat{b}^\dag\hat{j}
\end{equation}
which might differ from zero.

We also see that the result of the action of $\hat{\mathcal{T}}_\eta$ on $\ket{\psi _0}$ 
\begin{equation}
\hat{\mathcal{T}} _\eta \ket{\psi _0} = \left.\left. \left(\hat{\mathcal{T}}_x - \hat{\mathcal{T}}_y\right)\right|\psi _0 \right\rangle = \sum _{i=1}^N \sum _{a=N+1}^L (\tilde{\textbf{X}})_{i,a} \hat{a}^\dag\hat{i}\ket{\psi _0} 
\end{equation}
is not normalized since we have seen that
\begin{equation}
\braket{\psi _0 | \hat{\mathcal{T}}_\eta ^\dag \hat{\mathcal{T}}_\eta | \psi _0} = \braket{\psi _0 |\hat{\mathcal{T}}_x ^\dag \hat{\mathcal{T}}_x | \psi _0} = \vartheta _x \label{eq:not_norm_conserving}
\end{equation}
Note finally that such an application of $\hat{\mathcal{T}}_\eta$ on $\ket{\psi _0}$ might not be number-conserving. Indeed, we see that if we could write a quantum excited-state ansatz $\ket{\psi _{n,\eta}}$ from this application, the resulting density matrix, $\bm{\gamma}_{n,\eta}$, would have the following matrix elements
\begin{equation}
(\bm{\gamma}_{n,\eta})_{r,s} = \sum _{i,j=1}^N \sum _{a,b=N+1}^L (\tilde{\textbf{X}}\tmat)_{a,i}(\tilde{\textbf{X}})_{j,b}\underbrace{\braket{\psi _0| \hat{i}^\dag\hat{a} \hat{r}^\dag \hat{s}\hat{b}^\dag \hat{j}| \psi _0}}_{\lambda}
\end{equation}
leading to 
\begin{equation}
\bm{\gamma}_{n,\eta} = \left(\vartheta _x I_o - {\textbf{X}}{\textbf{X}}\tmat\right) \oplus {\textbf{X}}\tmat{\textbf{X}} \; \Longleftrightarrow \; \mathrm{tr}(\bm{\gamma}_{n,\eta}) = N \vartheta _x . \label{eq:not_number_conserving}
\end{equation}
As a conclusion, if $\hat{\mathcal{T}}_\eta$ had to be used as an approximation to the $\hat{\mathcal{T}}$ operator, it would not be nilpotent, nor norm-conserving and might not be number-conserving. It comes of course that such an approximation should not occurr, and that, as mentioned before, the $\hat{\mathcal{T}}_\eta$ operator should be seen as a corrected $\hat{\mathcal{T}}_x$ (i.e., the full CI truncated to single excitations) in the computation of matrix elements and transition energy. Indeed, while there would be no action of the deexcitation operator $\hat{\mathcal{T}}_y$ if $\hat{\mathcal{T}}_\eta$ was directly applied on $\ket{\psi _0}$, the reality of its action in the superoperators from Eqs. \eqref{eq:eom-diff} and \eqref{eq:transition_operators} has been seen directly when computing the matrix elements in this section and transition energy in Appendix \ref{app:EOM}.

\subsection{Switching off the particle/hole deexcitations}

If we now switch off the deexcitations, we come to another class of methods (denoted here by a ``$\zeta$" symbol) such as CIS or TDA. It has to be noted that while the following spinorbital-based derivation of the matrices structure is not directly applicable to the ADC method since the ADC ansatz consists in a linear combination of Intermediate States (ISs), the structure of the CIS/TDA density matrices in the space of singly-excited Slater determinants is transferrable to this method in the IS basis (see Appendix \ref{app:ADC}). The following derivations also hold for the TD-DFRT, ADC and LR--CC AMBWs reported in ref.\cite{crespo-otero_recent_2018}, but one has to keep in mind that these ansätze should be properly renormalized to yield the following matrices, and that in the LR--CC method, the Jacobian (\textbf{J}) is not Hermitian, so that there are left and right eigenvectors to it, on which the application of the following derivations might be performed for different purposes.\cite{crespo-otero_recent_2018}

The direct consequence of switching off the deexcitations when going from TDHF or TDDFT to CIS or TDA for instance, is that
\begin{equation}
\textbf{x}^\top _\zeta  \textbf{x}_\zeta^{\textcolor{white}{\dag}} - \underbrace{\textbf{y}^\top _\zeta  \textbf{y} _\zeta^{\textcolor{white}{\dag}}}_{0} = 1 \Longleftrightarrow \vartheta _x ^\zeta = 1.
\end{equation}
From this, we could postulate that there exists a nilpotent transition operator $\hat{\mathcal{T}}_\zeta$ leading to an excited-state ansatz
\begin{equation}\label{eq:Tzetapostulate}
\hat{\mathcal{T}} _\zeta \ket{\psi _0} = \sum _{i=1}^N \sum _{a=N+1}^L (\tilde{\textbf{X}}_\zeta)_{i,a} \hat{a}^\dag\hat{i}\ket{\psi _0} 
\end{equation}
with the excited state being orthogonal to $\ket{\psi _0}$, with $\hat{\mathcal{T}} _\zeta$ being norm- and number-conserving (we simply replace $\hat{\mathcal{T}}_\eta$ by $\hat{\mathcal{T}}_\zeta$ and $\vartheta _x$ by one in Eqs. \eqref{eq:not_norm_conserving} and \eqref{eq:not_number_conserving}). Unlike the $\eta$-methods, since we actually know the $\zeta$-ansatz from the methods themselves, we also know that the $\hat{\mathcal{T}}_\zeta$ we would insert in the EOM formulation of the transition energy and the EOM derivation of the transition and difference density matrix elements, is the actual $\ket{\psi _{n,\zeta}}\bra{\psi _0}$ operator. Note that, in this particular case, we could also derive the difference density matrix directly from the excited-state wave function
\begin{equation}
(\bm{\gamma}_{n,\zeta})_{r,s} = \sum _{i,j=1}^N \sum _{a,b=N+1}^L (\tilde{\textbf{X}}_\zeta^\top)_{a,i}(\tilde{\textbf{X}}_\zeta)_{j,b}\underbrace{\braket{\psi _0| \hat{i}^\dag\hat{a} \hat{r}^\dag \hat{s}\hat{b}^\dag \hat{j}| \psi _0}}_{\lambda}
\end{equation}
leading to 
\begin{equation}
\bm{\gamma}_{n,\zeta} =  \left(I_o - {\textbf{X}}_\zeta^{\textcolor{white}{\dag}}{\textbf{X}}_\zeta^\top\right) \oplus {\textbf{X}}^\top_\zeta{\textbf{X}}_\zeta ^{\textcolor{white}{\dag}} .
\end{equation}
Knowing that a single-reference ground state has a block-diagonal $\bm{\gamma}_0 = I_o \oplus 0_v$ state density matrix (see Appendix \ref{app:SD1-RDM}), it simply comes that
\begin{equation}
  \left(- {\textbf{X}}_\zeta^{\textcolor{white}{\dag}}{\textbf{X}}_\zeta^\top\right) \oplus {\textbf{X}}^\top_\zeta{\textbf{X}}_\zeta^{\textcolor{white}{\dag}}  = \bm{\gamma}^\Delta _\zeta \in \mathbb{R}^{L \times L}. \label{eq:DDM-zeta}
\end{equation}
For the transition density matrix, we have
\begin{equation}
(\bm{\gamma}^{0\rightarrow n} _\zeta)_{s,r} = \braket{\psi _0 | \hat{r}^\dag \hat{s}|\psi _{n,\zeta}} = \sum _{i=1}^N \sum _{a=N+1}^L \quad\textbf{x}_{ia} \!\!\!\!\!\!\!\!\!\!\!\!\underbrace{\braket{\psi _0 | \hat{r}^\dag \hat{s}\hat{a}^\dag\hat{i}|\psi _0}}_{\displaystyle{\delta_{rs}\delta_{ai} + \delta_{ri}n_r\delta_{sa}(1-n_s)}} = (\tilde{\textbf{X}})_{r,s} \, n_r (1-n_s)
\end{equation} 
leading to the nilpotent matrix
\begin{eqnarray}
  \left( 
 \begin{array}{cc}
0_o  & 0_{ov}   \\
\textbf{X}_\zeta^\top  &  0_v \\
 \end{array}\right) = \bm{\gamma}^{0 \rightarrow n} _\zeta \in \mathbb{R}^{L \times L} .\label{eq:TDM-zeta}
\end{eqnarray}
Expectedly, from Eqs. \eqref{eq:DDM-eta}, \eqref{eq:TDM-eta}, \eqref{eq:DDM-zeta}, and \eqref{eq:TDM-zeta}, we conclude that using directly the excited-state ansatz for $\zeta$-type methods to construct their transition and difference density matrices leads to the same results as if we used the EOM formalism, imposing the \textbf{Y} matrix elements (i.e., switching off the deexcitations) to vanish. As we can see from these four equations, the difference density matrix for these two types of methods has a similar shape, while the transition density matrix structure changes considerably when switching off the deexcitations (basically, the matrix becomes nilpotent). Since the difference and transition density matrices are involved in the qualitative and quantitative analyses of electronic transitions mentioned in the introduction, the alteration of the structure and content of the matrices, due to the choice of the excited-state calculation method, has an important impact on the expression of the descriptors that are used to describe the nature of the electronic excited states. 

\subsection{Summary}

\noindent Table \ref{table:working_equations} summarizes the results from this paper, and introduces the working equations\cite{crespo-otero_recent_2018,krause_implementation_2018} for the relevant excited-state calculation methods. For the $\eta$-methods we see the \textbf{A} (\textbf{B}) matrix appearing which, depending on the method, has a different content but invariably provides the single (de)excitations contributions to the transition energy. For the sake of generality, and in order to avoid confusion when dealing with the TD-DFRT AMBW, $\textbf{X}_\zeta$ in a nilpotent transition density matrix has been simply denoted by a \textbf{T}.

\begin{table}[h!]\label{table:working_equations}
\begin{center}
\includegraphics[scale=1]{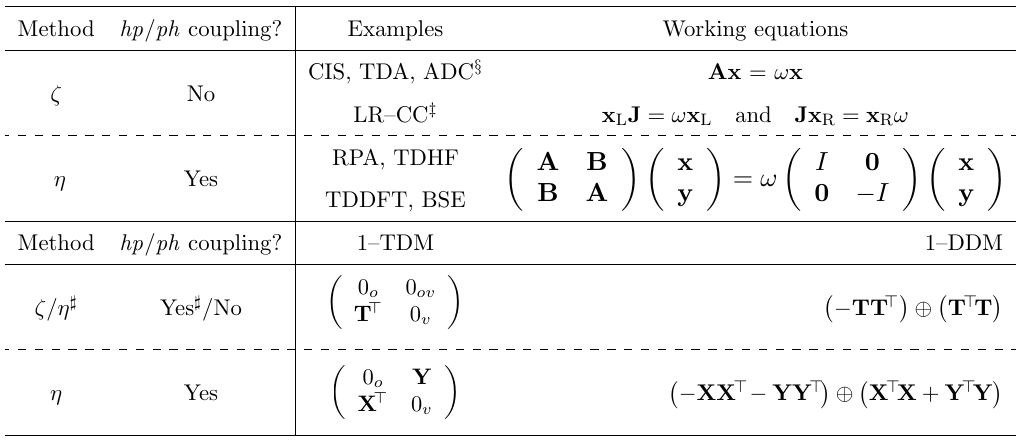}
\end{center}
\vspace*{-0.3cm}
\caption{Summary of the working equations and the one-particle transition density matrix (1--TDM) and difference density matrix (1--DDM) for the two types of excited-state calculation methods investigated in this contribution. Note that for the LR--CC method, we have the biorthogonality condition $\textbf{x}_\mathrm{L}^\top \textbf{x}_\mathrm{R}^{\textcolor{white}{\dag}} = 1$.  $^{\S}$The ADC AMBW;\cite{crespo-otero_recent_2018} $^{\ddag}$The LR--CC AMBW;\cite{crespo-otero_recent_2018} $^{\sharp}$Includes the TD--DFRT AMBW,\cite{crespo-otero_recent_2018} which produces matrices of the $\zeta$-type though the working equations are of $\eta$-type.}
\end{table}

\newpage

\section{Conclusion}

We reported a standalone derivation of the one-body density matrices of interest for the most common single-reference electronic excited-state calculation methods. We have first revisited in a comprehensive way a variant of Wick's theorem, an important tool in the second quantization derivation of density matrices in quantum chemistry. Afterward, we have introduced the basic equation-of-motion formalism, before applying it to two cases of excited-state calculation methods: those implying a partial correction to the full CI truncated to single excitations, and those solely implying a linear combination of singly-excited Slater determinants. The structure of the difference and transition density matrices was derived, discussed, and compared for the two types of methods. Finally, we gave a comment on the substitution of the transition operator in the first quantization equation-of-motion formalism by transition operators from the two types of methods discussed in this contribution.

\vspace*{-0.2cm}
\section*{Acknowledgements}

Drs Benjamin Lasorne, Matthieu Saubanère, and Emmanuel Fromager, are gratefully acknowledged for very fruitful discussions on the topic.

\section{Appendices}

\subsection{Basics of fermionic second quantization for this paper}\label{app:SQ}
In this appendix we recall the fundamentals of second quantization for fermions. For more details, see the introduction to second quantization in Ref.\cite{szabo1996szabo}. Note that in the following, the spinorbitals bases under consideration are assumed to orthogonal. This assumption is crucial for the second quantization properties used afterward.

Let's rewrite a one-determinant wavefunction value
\begin{eqnarray}
\psi(\textbf{s}_1,\ldots ,\textbf{s}_N) = \dfrac{1}{\sqrt{N !}}
\left|
\begin{array}{cccc}
{\varphi _1 (\textbf{s}_1)} & {\varphi _2 (\textbf{s}_1)} & \dots & {\varphi _N (\textbf{s}_1)}\\
{\varphi _1 (\textbf{s}_2)} & {\varphi _2 (\textbf{s}_2)} & \dots & {\varphi _N (\textbf{s}_2)}\\
\vdots & \vdots & \ddots & \vdots \\
{\varphi _1 (\textbf{s}_N)} & {\varphi _2 (\textbf{s}_N)} & \dots & {\varphi _N (\textbf{s}_N)}\\
\end{array}
\right| . \nonumber
\end{eqnarray}
The corresponding Fock state reads
\begin{equation}
 \ket{\psi} = \ket{\varphi _1 \varphi _2 \cdots \varphi _N}.
\end{equation}
We associate a {creation} operator $\hat{r}^\dag$ to a spinorbital $\varphi _r$. When acting on a one-determinant state, it creates an electron in $\varphi _r$
\begin{equation}
\nonumber \hat{a}_{r}^\dag \ket{\varphi _k \cdots \varphi_l} = \ket{\varphi _r \varphi _k \cdots \varphi _l}.
\end{equation}
The second quantization operators sequence order matters$\!\!$:
\begin{equation}
\hat{s}^\dag \hat{r}^\dag \ket{\varphi _u \cdots \varphi_v} = \ket{\varphi _s\varphi _r \varphi _u \cdots \varphi _v} = - \ket{\varphi _r\varphi _s \varphi _u \cdots \varphi _v} = - \, \hat{r}^\dag \hat{s}^\dag \ket{\varphi _u \cdots \varphi_v}.
\end{equation}
We therefore see that
\begin{equation}
 [ \hat{r}^\dag, \hat{s}^\dag ]_+ = \hat{0}. \label{eq:anticocreacrea}
\end{equation}
where the $[\cdot , \cdot]_+$ brackets denote the anticommutation superoperator:
\begin{equation}
[\hat{A},\hat{B}]_+ = \hat{A}\hat{B}+\hat{B}\hat{A}.
\end{equation}
From equation \eqref{eq:anticocreacrea} we find, setting $r=s$, that $\hat{r}^\dag\hat{r}^\dag = \hat{0}$.
Since a determinant with two identical columns (or lines) vanishes, we have
\begin{equation}\label{eq:SQzero1}
 r \, \in \, \{u, \cdots , v\} \; \Longleftrightarrow \; \hat{r}^\dag \ket{\varphi _u \cdots \varphi_v} = 0 
\end{equation}
which means that it is not possible to create an electron in $\varphi _r$ if that spinorbital is already occupied. 

One can create a Slater determinant using the creation operators
\begin{equation}\label{eq:refket}
\ket{\varphi _1 \varphi _2 \cdots \varphi _N} = \hat{\mathcal{C}}_N^\dag \ket{\,}
\end{equation}
with
\begin{equation}\label{eq:refketC}
\hat{\mathcal{C}}_N^\dag = \prod _{m=1}^N \hat{m}^\dag
\end{equation}
and $\ket{\,}$ the normalized (i.e., $\braket{\,|\,} = 1$) physical vacuum state.

The adjoint of any $\hat{r}^\dag$ creation operator is the $\hat{r}$ {annihilation} operator, and acts on a Slater determinant as
\begin{equation}
\hat{r} \ket{\varphi _r\varphi _k \cdots \varphi_l} = \ket{ \varphi _k \cdots \varphi _l}.
\end{equation}
The spinorbital sequence in the Slater determinant matters. Consider
\begin{equation}
\nonumber \hat{r}\ket{\varphi _u\varphi _v \cdots \varphi _r \cdots \varphi_w}.
\end{equation}
We first need to place the $\varphi _r$ spinorbital at the left of the Slater determinant before applying the second quantization operator $\hat{r}$. For this, we permute $\varphi _u$ and $\varphi _r$ (without forgetting the minus sign arising from the permutation of two matrix columns in a determinant)
\begin{align*}
\nonumber \hat{r}\ket{\varphi _u\varphi _v \cdots \varphi _r \cdots \varphi_w} &= - \hat{r}\ket{\varphi _r \varphi _v \cdots \varphi _u \cdots \varphi_w} \\ &= - \ket{ \varphi _v \cdots \varphi _u \cdots \varphi_w} \\ &= (-1)^{m(\sigma) + 1} \ket{ \varphi _u \varphi _v \cdots \varphi_w}
\end{align*}
where $m(\sigma)$ is the number of inversions necessary for bringing back $\varphi _u$ in its original position (the left of the Slater determinant), to which we add the original $\varphi _r$ permutation with $\varphi _u$ performed prior to the application of the annihilation operator $\hat{r}$ on the Slater determinant.
In general, for the natural ordering ($1,2,\cdots$) of the spinorbitals in the Slater determinant, we have
\begin{equation}
\nonumber \hat{r}\ket{\varphi _1\varphi _2 \cdots \varphi _r \cdots \varphi_N} = (-1)^{r-1} \ket{\varphi _1\varphi _2 \cdots \varphi_N}
\end{equation}
with one permutation to place $\varphi _r$ at the left of the Slater determinant, and $r-2$ for replacing $\varphi _1$ in its place after the annihilation of the electron in $\varphi _r$. 

As for the creation operators, we also have
\begin{equation}
 [ \hat{r} , \hat{s}  ]_+ = \hat{0} \quad \mathrm{and}\quad \hat{r}\hat{r} = \hat{0}
\end{equation}
and
\begin{equation}\label{eq:SQzero2}
r \, \notin \, \{u, \cdots , v\} \; \Longleftrightarrow \; \hat{r}  \ket{\varphi _u \cdots \varphi_v} = 0 .
\end{equation}
One can also demonstrate the following property
\begin{equation}
\left[\hat{r}^\dag,\hat{s} \right]_+  = {\delta}_{rs} = \left[\hat{r},\hat{s}^\dag \right]_+.\label{eq:anticommutationdeltars}
\end{equation}
Indeed, we see that, independently from the occupation status of $\varphi _r$ and $\varphi _s$ in $\ket{\varphi _u \cdots \varphi _v}$,
\begin{equation}
[\hat{r},\hat{s}^\dag]_+\ket{\varphi _u \cdots \varphi _v} = \ket{\varphi _u \cdots \varphi _v} \Longleftrightarrow (r=s).
\end{equation}
and
\begin{equation}
[\hat{r},\hat{s}^\dag]_+\ket{\varphi _u \cdots \varphi _v} = 0 \Longleftrightarrow (r \neq s) .
\end{equation}
What is true for any $[\hat{r},\hat{s}^\dag]_+$ in these conditions will remain true for any $[\hat{r}^\dag,\hat{s}]_+$ in the same conditions, so we deduce equation \eqref{eq:anticommutationdeltars}.

\noindent Note also that, knowing that $\left( \hat{r}^\dag \right)^\dag = \hat{r}$, we find that in the dual space, a creation operator becomes an annihilation operator and \textit{vice versa}.

Considering the fact that for any two operators $\hat{A}$ and $\hat{B}$ we have 
\begin{equation}
\left(\hat{{A}}\hat{{B}}\right)^{\dag} = \hat{{B}}^\dag \hat{{A}}^\dag,
\end{equation} 
we can deduce that, in the dual space,
\begin{equation}
\bra{\varphi _N \cdots \varphi _2 \varphi _1} = \bra{\,} \hat{\mathcal{C}}_N  \label{eq:refdual}
\end{equation}
Finally, if $1 < i < N$ and $a > N$, a singly-excited Slater configuration reads, in second quantization,
\begin{equation}
\nonumber \ket{\psi _i ^a} = \hat{a}^\dag \hat{i} \ket{\varphi _1 \cdots \varphi _i \cdots \varphi _N}.
\end{equation}
This naturally extends to $1 \leq i \leq N$.
\subsection{Decomposition of chains of second quantization operators expectation values}\label{app:wick}
\noindent In this appendix we will see how one can use the time-independent Wick's theorem to simplify the evaluation of the matrix elements of this contribution. For more details on the topic, we recommend Refs.\cite{shavitt_many-body_2009,lindgren2012atomic,bogolyubov1980introduction} and references therein.
\subsubsection{``{Augmented}" second quantization operators}
\noindent Similar to the behavior of second quantization operators relatively to the physical vacuum,
\begin{equation}
\hat{r}\ket{\,} = 0 \quad \mathrm{and}\quad  \bra{\,}\hat{r}^\dag = 0 \label{eq:pre-lemma1}
\end{equation}
we build two types of second quantization operators that will have the exact same behavior but relatively to a single-determinant (SD) reference state $\ket{\psi _0}$, namely
\begin{equation}
\hat{p}_r\ket{\psi _0} = 0 \quad \mathrm{and}\quad  \bra{\psi _0}\hat{p}^\dag_r = 0. \label{eq:pre-lemma2}
\end{equation}
For any spinorbital $\varphi _r$ in the canonical space, we can create two ``{augmented}" second quantization operators:
\begin{align}\label{eq:pseudo1}
\hat{p}_r^\dag &= \hat{r}\times n_r + \hat{r}^\dag \times  (1-n_r),\\
\hat{p}_r &= \hat{r}\times (1-n_r)+\hat{r}^\dag \times n_r.\label{eq:pseudo2}
\end{align}
where $n_r$ is the occupation number of the $\varphi _r$ spinorbital in the SD reference state $\ket{\psi _0}$. It takes the value $1$ if the spinorbital is occupied, and $0$ if it is unoccupied. This number is sometimes written $n_r(\psi _0)$, but here we will use the shorter $n_r$ since any occupation number in this appendix will be given relatively to the SD reference state $\ket{\psi _0}$.

As an illustration, we see that if $n_i= 1$ and $n_a  = 0$,
\begin{align*}
\hat{p}_i &=  \hat{i}^\dag \\ &= \hat{i}\times (1-1)+\hat{i}^\dag \times 1, \\ \hat{p}^\dag _i &=  \hat{i} \\&= \hat{i}\times 1 + \hat{i}^\dag \times  (1-1),
\end{align*}
and
\begin{align*}
\hat{p}_a &= \hat{a} \\ &=\hat{a}\times (1-0)+\hat{a}^\dag \times 0, \\ \hat{p}^\dag _a &= \hat{a}^\dag \\ &= \hat{a}\times 0 + \hat{a}^\dag \times  (1-0).
\end{align*}
\subsubsection{Normal-ordered chain of operators}
\noindent A normal-ordered chain of creation/annihilation operators relatively to the physical vacuum reads
\begin{equation}\label{eq:normal-vacuum}
n[\hat{Q}_1\hat{Q}_2\hat{Q}_3\cdots] := (-1)^{m(\sigma)} \hat{c}_1^\dag \hat{c}_2^\dag\cdots \hat{c}_2\hat{c}_1,
\end{equation}
where all the creation operators are at the left of all the annihilation operators. For example, for the $\hat{u}^\dag\hat{v}\hat{w}^\dag\hat{x}\hat{y}^\dag\hat{z}$ chain, we have
\begin{equation}
n[\hat{u}^\dag\hat{v}\hat{w}^\dag\hat{x}\hat{y}^\dag\hat{z}] = (-1)^{m(\sigma)} \hat{u}^\dag\hat{w}^\dag\hat{y}^\dag\hat{v}\hat{x}\hat{z}
\end{equation}
where $m(\sigma) = 3$ is the number of inversions in the $\sigma$ permutation 
\begin{equation} 
\sigma :=
\begin{pmatrix}
\hat{u}^\dag\hat{v}\hat{w}^\dag\hat{x}\hat{y}^\dag\hat{z} \\
\hat{u}^\dag\hat{w}^\dag\hat{y}^\dag\hat{v}\hat{x}\hat{z}
\end{pmatrix}.
\end{equation}
A normal-ordered chain of creation/annihilation operators with respect to the SD reference state $\ket{\psi _0}$ reads
\begin{equation}\label{eq:normal-fermi}
N[\hat{Q}_1\hat{Q}_2\hat{Q}_3\cdots] := (-1)^{m(\sigma)} \hat{p}_p^\dag \hat{p}_q^\dag \cdots \hat{p}_u \hat{p}_v
\end{equation}
Note the lower-case $n$ for the permutation relative to the physical vacuum, and upper-case $N$ for the permutation relative to $\ket{\psi _0}$. We see that the $n$ permutation is performed solely based on the creation/annihilation character of the operators, while the $N$ permutation takes into account both the creation/annihilation character of the operators and the occupation numbers in the SD reference state $\ket{\psi _0}$. For illustrating the difference between the two permutations, we take the $\hat{i}^\dag\hat{a}\hat{b}^\dag\hat{j}$ chain, for which we give $n_i(\psi _0) = n_j(\psi _0) = 1$ and $n_a (\psi _0) = n_b(\psi _0) = 0$. Therefore,
\begin{equation}
n[\hat{i}^\dag\hat{a}\hat{b}^\dag\hat{j}] = (-1)^1\, \hat{i}^\dag \hat{b}^\dag \hat{a}\hat{j}, \quad N[\hat{i}^\dag\hat{a}\hat{b}^\dag\hat{j}] = (-1)^4 \, \hat{b}^\dag\hat{j}\hat{i}^\dag\hat{a}.
\end{equation}
Using the results in equations (\ref{eq:pre-lemma1}) and (\ref{eq:pre-lemma2}), we have the following lemma:
\begin{lemma}\label{lem:expnormzero}
The physical vacuum/SD reference state expectation value of a normal-ordered chain of second quantization operators is  zero.
\end{lemma}
\noindent which also reads
\begin{equation}
\braket{\, | n[\hat{Q}_1\hat{Q}_2\cdots]|\,} = 0, \quad \braket{\psi _0 |N[\hat{Q}_1\hat{Q}_2\cdots]|\psi _0} = 0.
\end{equation}

\subsubsection{Contractions}
\noindent We define the contractions of any two second quantization operators ($\hat{Q}_1$ and $\hat{Q}_2$) as
\begin{equation}\label{eq:contractions}
\contraction{}{\hat{Q}}{_1}{\hat{Q}}
\bcontraction{\hat{Q}_1\hat{Q}_2 := \hat{Q}_1\hat{Q}_2 - n[\hat{Q}_1\hat{Q}_2], \quad\,}{\hat{Q}}{_1}{\hat{Q}}
\hat{Q}_1\hat{Q}_2 := \hat{Q}_1\hat{Q}_2 - N[\hat{Q}_1\hat{Q}_2], \quad \hat{Q}_1\hat{Q}_2 := \hat{Q}_1\hat{Q}_2 - n[\hat{Q}_1\hat{Q}_2],
\end{equation}
where again $n$ is the product in normal order relative to the physical vacuum, and $N$ is the product in normal order relative to the SD reference state $\ket{\psi _0}$. Note that this definition extends to the case where there is a product of second quantization operators between the contracted ones
\begin{equation}\label{eq:contractjump}
\contraction{}{\hat{Q}}{\left(\hat{Q}_1\cdots\hat{Q}_n\right)}{\hat{Q}}
\bcontraction[1.5ex]{\hat{Q}\left(\hat{Q}_1\cdots\hat{Q}_n\right)\hat{Q}' = (-1)^n \hat{Q}\hat{Q}'\left(\hat{Q}_1\cdots\hat{Q}_n\right), \quad \!}{\hat{Q}}{\left(\hat{Q}_1\cdots\hat{Q}_n\right)\!}{\hat{Q}}
\contraction{\hat{Q}\left(\hat{Q}_1\cdots\hat{Q}_n\right)\hat{Q}' = (-1)^n }{\hat{Q}}{}{\hat{Q}}
\bcontraction[1.5ex]{\hat{Q}\left(\hat{Q}_1\cdots\hat{Q}_n\right)\hat{Q}' = (-1)^n \hat{Q}\hat{Q}'\left(\hat{Q}_1\cdots\hat{Q}_n\right), \quad \hat{Q}\left(\hat{Q}_1\cdots\hat{Q}_n\right)\hat{Q}' = (-1)^n \!}{\hat{Q}}{\!}{\hat{Q}}
\hat{Q}\left(\hat{Q}_1\cdots\hat{Q}_n\right)\hat{Q}' = (-1)^n \hat{Q}\hat{Q}'\left(\hat{Q}_1\cdots\hat{Q}_n\right), \quad \hat{Q}\left(\hat{Q}_1\cdots\hat{Q}_n\right)\hat{Q}' = (-1)^n \hat{Q}\hat{Q}'\left(\hat{Q}_1\cdots\hat{Q}_n\right).
\end{equation}

$\;$

\noindent \textbf{Contractions relative to the physical vacuum}

$\;$

\noindent We have, according to equations \eqref{eq:normal-vacuum} and \eqref{eq:contractions},
\begin{equation}\label{eq:normalvacuumzero}
\bcontraction{\!}{\hat{r}}{^\dag\,}{\hat{s}}
\bcontraction{\hat{r}^\dag\hat{s} = 0, \quad \!}{\hat{r}}{^\dag\,}{\hat{s}}
\bcontraction{\hat{r}^\dag\hat{s} = 0, \quad \hat{r}^\dag\hat{s}^\dag = 0, \quad \!}{\hat{r}}{\,}{\hat{s}}
\hat{r}^\dag\hat{s} = 0, \quad \hat{r}^\dag\hat{s}^\dag = 0, \quad \hat{r}\hat{s} = 0,
\end{equation}
and, according to the anticommutation rule for fermionic second quantization operators in equation \eqref{eq:anticommutationdeltars},
\begin{equation}\label{eq:outnormalvacuum}
\bcontraction{\!}{\hat{s}}{\,}{\hat{r}}
\hat{s}\hat{r}^\dag = \hat{s}\hat{r}^\dag - n[\hat{s}\hat{r}^\dag] = \hat{s}\hat{r}^\dag -\left(- \hat{r}^\dag\hat{s}\right) =  {\delta}_{rs} .
\end{equation}

$\;$

\noindent \textbf{Contractions with respect to the SD reference state} $\ket{\psi _0}$

$\;$

\noindent In this case we must not only care about the creation/annihilation nature of the operators, but distinguish the operators relative to spinorbitals that are occupied and unoccupied in $\ket{\psi _0}$. 
Combining equation \eqref{eq:contractions} to lemma \ref{lem:expnormzero}, we find 
\begin{lemma}\label{lem:contraction}
The contraction $($relative to a SD reference state $\ket{\psi _0})$ of a product of two second quantization operators is equal to the expectation value of this product, relative to that reference state.
\end{lemma}
\noindent This lemma also reads
\begin{equation}
\contraction{}{\hat{Q}}{_1}{\hat{Q}}%
\hat{Q}_1\hat{Q}_2 = \left\langle \psi _0 \left|\hat{Q}_1\hat{Q}_2\right|\psi _0\right\rangle.
\end{equation}
The four possibilities are
\begin{align}
\contraction{}{\hat{r}}{^\dag}{\hat{s}}
\hat{r}^\dag\hat{s} &= \braket{\psi _0 | \hat{r}^\dag\hat{s}|\psi _0} =  \delta_{rs}n_rn_s,
\\
\contraction{}{\hat{r}}{}{\hat{s}}
\hat{r}\hat{s}^\dag &= \braket{\psi _0 | \hat{r}\hat{s}^\dag|\psi _0} = \delta_{rs}(1-n_r)(1-n_s),
\\
\contraction{}{\hat{r}}{^\dag}{\hat{s}}
\hat{r}^\dag\hat{s}^\dag &= \braket{\psi _0 | \hat{r}^\dag\hat{s}^\dag|\psi _0} = \delta_{rs}n_r(1-n_s) = 0 ,
\\
\contraction{}{\hat{r}}{}{\hat{s}}
\hat{r}\hat{s} &= \braket{\psi _0 | \hat{r}\hat{s}|\psi _0} = \delta_{rs}(1-n_r)n_s = 0 .
\end{align}
\subsubsection{Normal-ordered chains with contractions}
We can define contracted normal-ordered chains of second quantization operators as
\begin{equation}\label{eq:contracteduppern}
\contraction[1ex]{N[\hat{Q}_1\hat{Q}_2\hat{Q}_3\cdots}{\hat{Q}}{_4\cdots \hat{Q}_5 \cdots}{\hat{Q}}%
\contraction[2ex]{N[\hat{Q}_1\hat{Q}_2\hat{Q}_3\cdots\hat{Q}_4\cdots}{ \hat{Q}}{_5 \cdots\hat{Q}_6\cdots}{\hat{Q}}
\contraction[1ex]{N[\hat{Q}_1\hat{Q}_2\hat{Q}_3\cdots \hat{Q}_4 \cdots \hat{Q}_5 \cdots \hat{Q}_6 \cdots \hat{Q}_7\cdots] := (-1)^{m(\sigma)}}{\hat{Q}}{_4}{\hat{Q}}
\contraction[1ex]{N[\hat{Q}_1\hat{Q}_2\hat{Q}_3\cdots \hat{Q}_4 \cdots \hat{Q}_5 \cdots \hat{Q}_6 \cdots \hat{Q}_7\cdots] :=(-1)^{m(\sigma)} \hat{Q}_4\hat{Q}_6 \, }{\hat{Q}}{_5}{\hat{Q}}
N[\hat{Q}_1\hat{Q}_2\hat{Q}_3\cdots \hat{Q}_4 \cdots \hat{Q}_5 \cdots \hat{Q}_6 \cdots \hat{Q}_7\cdots] :=(-1)^{m(\sigma)} \hat{Q}_4\hat{Q}_6 \, \hat{Q}_5\hat{Q}_7 \cdots N[\hat{Q}_1\hat{Q}_2\hat{Q}_3\cdots]
%
%
%
%
\end{equation}
relatively to the $\ket{\psi _0}$ SD reference state and
\begin{equation}\label{eq:contractedlowern}
\bcontraction[1ex]{n[\hat{Q}_1\hat{Q}_2\hat{Q}_3\cdots}{\!\hat{Q}}{_4\cdots \hat{Q}_5 \cdots}{\hat{Q}}%
\bcontraction[2ex]{n[\hat{Q}_1\hat{Q}_2\hat{Q}_3\cdots\hat{Q}_4\cdots}{\!\hat{Q}}{_5 \cdots\hat{Q}_6\cdots}{\hat{Q}}
\bcontraction[1ex]{n[\hat{Q}_1\hat{Q}_2\hat{Q}_3\cdots \hat{Q}_4 \cdots \hat{Q}_5 \cdots \hat{Q}_6 \cdots \hat{Q}_7\cdots] := (-1)^{m(\sigma)}}{\!\hat{Q}}{_4}{\hat{Q}}
\bcontraction[1ex]{n[\hat{Q}_1\hat{Q}_2\hat{Q}_3\cdots \hat{Q}_4 \cdots \hat{Q}_5 \cdots \hat{Q}_6 \cdots \hat{Q}_7\cdots] :=(-1)^{m(\sigma)} \hat{Q}_4\hat{Q}_6 \, }{\!\hat{Q}}{_5}{\hat{Q}}
n[\hat{Q}_1\hat{Q}_2\hat{Q}_3\cdots \hat{Q}_4 \cdots \hat{Q}_5 \cdots \hat{Q}_6 \cdots \hat{Q}_7\cdots] :=(-1)^{m(\sigma)} \hat{Q}_4\hat{Q}_6 \, \hat{Q}_5\hat{Q}_7 \cdots n[\hat{Q}_1\hat{Q}_2\hat{Q}_3\cdots]
%
%
%
%
\end{equation}
relatively to the physical vacuum. In both cases, $m(\sigma)$ is the number of inversions in the $\sigma$ permutation 
\begin{equation} 
\sigma :=
\begin{pmatrix}
\hat{Q}_1\hat{Q}_2\hat{Q}_3\cdots \hat{Q}_4\cdots \hat{Q}_5\cdots\hat{Q}_6\cdots\hat{Q}_7\cdots \\
\hat{Q}_4\hat{Q}_6\hat{Q}_5 \hat{Q}_7\cdots \hat{Q}_1\hat{Q}_2\hat{Q}_3\cdots 
\end{pmatrix}.
\end{equation}
From Lemma \ref{lem:expnormzero} and definitions (\ref{eq:contracteduppern}) and (\ref{eq:contractedlowern}), we find
\begin{lemma}\label{lem:expnormfully}
The single-reference expectation value of a contracted normal-ordered chain of second quantization operators is equal to zero unless the product is fully contracted.
\end{lemma}
\noindent The number of possible full contractions of a product of $2n$ operators is $\displaystyle (2n-1)!! = \dfrac{(2n)!}{n! \, 2^n}$, as derived in Appendix \ref{app:NQ}.
\subsubsection{Wick's theorem}
\noindent We will bring here a tool extensively used in quantum field theory: the \textit{time-independent Wick's theorem}. In particular, its adaptation to normal-ordered products relative to a SD reference state, and the corollary to it will be of great interest to us in this contribution. The theorem itself reads
\begin{theorem}\label{th:wick}
Any chain of second quantization operators can be rewritten as the corresponding normal-ordered product plus the sum of all of the possible contracted normal-ordered products.
\end{theorem}
\noindent This theorem also reads

$\;$

\begin{equation}\label{eq:wickphysvac}\hspace*{-15cm}\begin{tikzpicture}
$\displaystyle \hat{Q}_1\hat{Q}_2\hat{Q}_3\hat{Q}_4 \cdots  = n[\hat{Q}_1\hat{Q}_2  \hat{Q}_3\hat{Q}_4 \cdots]
+ \!\!\!\!\! \sum _{\substack{ \mathrm{all} \\ \mathrm{contractions}}} \!\!\! \!\!n[\hat{Q}_1\hat{Q}_2  \hat{Q}_3\hat{Q}_4 \cdots]$
\draw [-,join=round] (-1.45,-.25)--(-2.45,-.25)--(-2.45,-.1);
\draw [-,join=round] (-0.975,-.4)--(-1.98,-.4)--(-1.98,-.1);
\end{tikzpicture}\end{equation}

$\;$

\noindent Due to the construction of the ``augmented" second quantization operators and the identical structure of the tools elaborated relatively to the physical vacuum or the SD reference state $\ket{\psi _0}$, the theorem is fully transferrable to normal-ordered chains of second quantization operators relative to the SD reference state $\ket{\psi _0}$,

\begin{equation}\label{eq:wickfermivac}\hspace*{-15cm}\begin{tikzpicture}
$\displaystyle \hat{Q}_1\hat{Q}_2\hat{Q}_3\hat{Q}_4 \cdots  = N[\hat{Q}_1\hat{Q}_2  \hat{Q}_3\hat{Q}_4 \cdots]
+ \!\!\!\!\! \sum _{\substack{ \mathrm{all} \\ \mathrm{contractions}}} \!\!\! \!\!N[\hat{Q}_1\hat{Q}_2  \hat{Q}_3\hat{Q}_4 \cdots]$
\draw [-,join=round] (-1.45,.55)--(-2.385,.55)--(-2.385,.4);
\draw [-,join=round] (-0.975,.7)--(-1.915,.7)--(-1.915,.4);
\end{tikzpicture}\end{equation}

$\;$

$\;$

$\;$

\noindent Note that in the two formulations of the theorem, the sum holds for all the singly-, doubly-(,...)contracted normal-ordered products.

According to what precedes, the demonstration of the theorem is identical using any of the two references (physical vacuum or SD reference state). Here we will use the SD reference state $\ket{\psi _0}$ for demonstrating time-independent Wick's theorem. 

$\;$

\noindent \textbf{Proof by induction}

$\;$

\noindent We need to prove that if the theorem is valid for any product of $n$ operators (induction hypothesis, with $n \geq 1$), it holds for $n+1$ operators (step case). If the last statement is true, and if the theorem holds for the case where $n=1$ (base case), then we have that the theorem holds for any $n$.

$\;$

\noindent \textit{Induction hypothesis}: equation \eqref{eq:wickfermivac} is assumed to hold for any chain $\hat{Q} = \hat{Q}_1\hat{Q}_2\cdots \hat{Q}_n$ of second quantization operators. The sum in the right-hand side of equation \eqref{eq:wickfermivac} will be written $S[\hat{Q}]$ in what follows.

$\;$

\noindent \textit{Step case}: we need to show that equation \eqref{eq:wickfermivac} holds for any chain $\hat{P} = \hat{Q}\hat{Q}_{n+1}$, with $\hat{Q}_{n+1}$ being any (creation or annihilation) operator. The theorem for $\hat{P}$ reads
\begin{equation}\label{eq:wickP}
\hat{P} = N[\hat{P}]+S[\hat{P}].
\end{equation}
If the theorem is true,
\begin{equation}\label{eq:PQQnp1}
\hat{P} = \hat{Q}\hat{Q}_{n+1} = N[\hat{Q}]\hat{Q}_{n+1} + S[\hat{Q}]\hat{Q}_{n+1}
\end{equation}
\textbf{Case A} -- $\hat{Q}_{n+1}$ is an annihilation operator. Then,
\begin{equation}
N[\hat{Q}]\hat{Q}_{n+1} = N[\hat{P}], \qquad S[\hat{Q}]\hat{Q}_{n+1} = N[ S[\hat{Q}]\hat{Q}_{n+1}] =  S[\hat{P}],
\end{equation}
the last equality beign true due to the fact that any contraction with $\hat{Q}_{n+1}$ vanishes if $\hat{Q}_{n+1}$ is placed at the right of $\hat{Q}$. In these conditions, we find equation \eqref{eq:wickP}, so the theorem holds in this case.

$\;$

\noindent \textbf{Case B} -- $\hat{Q}_{n+1}$ is a creation operator. Then, using equations \eqref{eq:contractions} and \eqref{eq:contractjump}, we have that
\begin{equation}
\contraction{\hat{P} = \hat{Q}_1\hat{Q}_2\cdots\hat{Q}_{n}\hat{Q}_{n+1} = - \underbrace{\hat{Q}_1\hat{Q}_2\cdots \hat{Q}_{n+1}\hat{Q}_n}_{\displaystyle x} + \hat{Q}_1\hat{Q}_2\cdots }{\hat{Q}}{_{n}}{\hat{Q}}
\hat{P} = \hat{Q}_1\hat{Q}_2\cdots\hat{Q}_{n}\hat{Q}_{n+1} = - \underbrace{\hat{Q}_1\hat{Q}_2\cdots \hat{Q}_{n+1}\hat{Q}_n}_{\displaystyle x} + \hat{Q}_1\hat{Q}_2\cdots \hat{Q}_{n}\hat{Q}_{n+1},
\end{equation}
with
\begin{equation}
\contraction{x = - \hat{Q}_1\hat{Q}_2\cdots \hat{Q}_{n+1}\hat{Q}
_{n-1}\hat{Q}_n + \hat{Q}_1\hat{Q}_2\cdots}{\hat{Q}}{_{n-1}}{\hat{Q}}
x = - \hat{Q}_1\hat{Q}_2\cdots \hat{Q}_{n+1}\hat{Q}
_{n-1}\hat{Q}_n + \hat{Q}_1\hat{Q}_2\cdots \hat{Q}_{n-1}\hat{Q}_{n+1}\hat{Q}_{n},
\end{equation}
i.e., using \eqref{eq:contractjump},
\begin{equation}
\contraction{x= - \hat{Q}_1\hat{Q}_2\cdots \hat{Q}_{n+1}\hat{Q}
_{n-1}\hat{Q}_n + (-1)^1\hat{Q}_1\hat{Q}_2\cdots}{\hat{Q}}{_{n-1}\hat{Q}_{n}}{\hat{Q}}
x= - \hat{Q}_1\hat{Q}_2\cdots \hat{Q}_{n+1}\hat{Q}
_{n-1}\hat{Q}_n + (-1)^1\hat{Q}_1\hat{Q}_2\cdots \hat{Q}_{n-1}\hat{Q}_{n}\hat{Q}_{n+1},
\end{equation}
so that 
\begin{equation}
\contraction{\hat{P} = \underbrace{\hat{Q}_1\hat{Q}_2\cdots \hat{Q}_{n+1}\hat{Q}
_{n-1}\hat{Q}_n}_{\displaystyle y} + \hat{Q}_1\hat{Q}_2\cdots }{\hat{Q}}{_{n}}{\hat{Q}}
\contraction{\hat{P} = \underbrace{\hat{Q}_1\hat{Q}_2\cdots \hat{Q}_{n+1}\hat{Q}
_{n-1}\hat{Q}_n}_{\displaystyle y} + \hat{Q}_1\hat{Q}_2\cdots \hat{Q}_{n}\hat{Q}_{n+1} + \hat{Q}_1\hat{Q}_2\cdots }{\hat{Q}}{_{n-1}\hat{Q}_{n}}{\hat{Q}}
\hat{P} = \underbrace{\hat{Q}_1\hat{Q}_2\cdots \hat{Q}_{n+1}\hat{Q}
_{n-1}\hat{Q}_n}_{\displaystyle y} + \hat{Q}_1\hat{Q}_2\cdots \hat{Q}_{n}\hat{Q}_{n+1} + \hat{Q}_1\hat{Q}_2\cdots \hat{Q}_{n-1}\hat{Q}_{n}\hat{Q}_{n+1}.
\end{equation}
The following step consists in developing $y$ as we developed $x$. We then procede likewise until $\hat{Q}_{n+1}$ is at the left of $\hat{Q}$, and find
\begin{equation}
\contraction{\hat{P} = (-1)^n \hat{Q}_{n+1}\hat{Q} + \sum _{1 \leq i \leq n} \hat{Q}_{1}\hat{Q}_2\cdots }{\hat{Q}}{_i\cdots\hat{Q}_n}{\hat{Q}}
\hat{P} = (-1)^n \hat{Q}_{n+1}\hat{Q} + \sum _{1 \leq i \leq n} \hat{Q}_{1}\hat{Q}_2\cdots \hat{Q}_i\cdots\hat{Q}_n\hat{Q}_{n+1}
\end{equation}
In the special case $\hat{Q} = N[\hat{Q}]$, we have
\begin{equation}
\contraction{ N[\hat{Q}]\hat{Q}_{n+1} = (-1)^n\hat{Q}_{n+1}N[\hat{Q}] + }{N}{[\hat{Q}]}{\hat{Q}}
 N[\hat{Q}]\hat{Q}_{n+1} = (-1)^n\hat{Q}_{n+1}N[\hat{Q}] + N[\hat{Q}]\hat{Q}_{n+1}.
\end{equation}
Since
\begin{equation}
\hat{Q}_{n+1}N[\hat{Q}] = N [\hat{Q}_{n+1}\hat{Q}] = (-1)^n N[\hat{Q}\hat{Q}_{n+1}],
\end{equation}
we have
\begin{equation}\label{eq:NQQnp1}
\contraction{N[\hat{Q}]\hat{Q}_{n+1} = N[\hat{Q}\hat{Q}_{n+1}] + }{N}{[\hat{Q}]}{\hat{Q}}
N[\hat{Q}]\hat{Q}_{n+1} = N[\hat{Q}\hat{Q}_{n+1}] + N[\hat{Q}]\hat{Q}_{n+1}.
\end{equation}
We can apply this procedure again to find
\begin{equation}\label{eq:SQQnp1}
\contraction{S[\hat{Q}]\hat{Q}_{n+1} = N[S[\hat{Q}]\hat{Q}_{n+1}] + }{S}{[\hat{Q}]}{\hat{Q}}
S[\hat{Q}]\hat{Q}_{n+1} = N[S[\hat{Q}]\hat{Q}_{n+1}] + S[\hat{Q}]\hat{Q}_{n+1}.
\end{equation}
Adding equations \eqref{eq:NQQnp1} to \eqref{eq:SQQnp1}, and injecting the result in equation \eqref{eq:PQQnp1}, we find that
\begin{equation}
\contraction{\hat{P} = N[\hat{Q}\hat{Q}_{n+1}] + }{N}{[\hat{Q}]}{\hat{Q}}%
\contraction{\hat{P} = N[\hat{Q}\hat{Q}_{n+1}] + N[\hat{Q}]\hat{Q}_{n+1} + N[S[\hat{Q}]\hat{Q}_{n+1}] + }{S}{[\hat{Q}]}{\hat{Q}}
\hat{P} = N[\hat{Q}\hat{Q}_{n+1}] + \underbrace{N[\hat{Q}]\hat{Q}_{n+1} + N[S[\hat{Q}]\hat{Q}_{n+1}] + S[\hat{Q}]\hat{Q}_{n+1}}_{\displaystyle S[\hat{Q}\hat{Q}_{n+1}]},
\end{equation}
i.e., we find equation \eqref{eq:wickP}, so the theorem holds for $n+1$ operators. Since it naturally holds for $n=1$, we conclude that it holds for any $n$. \qed

$\;$

Combining lemma \ref{lem:expnormzero} and lemma \ref{lem:contraction} with theorem \ref{th:wick}, we find
\begin{corollary}\label{cor:exp}
The single-reference expectation value of a chain of $2M$ fermionic second quantization operators can be decomposed into a sum of products of $M$ two-operator expectation values, each product being affected by a sign.
\end{corollary}
\noindent Corollary \ref{cor:exp} also reads
\begin{equation}
\left\langle \psi _0 \left |\, \prod _{k=1}^{2M} \hat{Q}_k \, \right|\psi _0\right\rangle = \sum _{h = 1} ^{(2M-1)!!} (-1)^{m_h} \prod _{e=1}^{M}  {\left\langle\psi _0 \left| {}^{[e]}\hat{q} _{1,h} {}^{[e]}\hat{q}_{2,h} \right| \psi _0\right\rangle}
\end{equation}
where the $\hat{Q}_k$ and the ${}^{[e]}\hat{q} _{i,h}$ operators can be either creation or annihilation operators, and $m_h$ is the resulting signature of the $h^{\mathrm{th}}$ full contraction.

$\;$

\noindent \textbf{Why are we using the ``augmented" second quantization operators?}

$\;$

\noindent This choice is motivated by the fact that, according to equations \eqref{eq:refket}, \eqref{eq:refketC}, and \eqref{eq:refdual}, we have, for any product $\hat{Q}$ of $2M$ operators (in the text we deal with even numbers of operators) relatively to an $N-$electron one-determinant reference state $\ket{\psi _0}$,
\begin{equation}
\left\langle \psi _0 \left |\, \hat{Q}\, \right|\psi _0\right\rangle = \left\langle \, \left| \hat{\mathcal{C}}\hat{Q}\hat{\mathcal{C}}^\dag \right| \, \right\rangle,
\end{equation}
with $2(M+N)$ operators in $\hat{\mathcal{C}}\hat{Q}\hat{\mathcal{C}}^\dag$, hence leading to 
\begin{equation}
(2M+2N-1)!! = \dfrac{(2M+2N)!}{(M+N)!\,2^M2^N}
\end{equation}
possible full contractions relative to the physical vacuum rather than $(2M-1)!!$ if we use the SD reference state $\ket{\psi _0}$. Note finally that in the case of an integral as $\mathscr{I}_M$ in equation \eqref{eq:I_M}, there are only $M$ non-vanishing contractions that are possible for a fully-contracted product, leading to $M!$ possible full contractions, hence solely $M!$ terms in the expectation value decomposition.

\subsection{Determination of $\mathscr{N}_Q$}\label{app:NQ}
When pairing one of the $Q$ $(=2M)$ operators with another one from the $Q-$chain, one can choose among the $Q-1$ (1 being itself) remaining operators. When constituting a second pair of operators, the choice is reduced to $Q - 1 -2$, where 2 is pointing the first pair already constituted, and 1 is pointing the operator we want to pair with another one. For a third pairing we have the choice among the $Q$ operators reduced by the number of already paired operators (4, i.e., 2 pairs), and the operator we want to pair, so the number of possibilities is reduced to $Q-1 -2 \times 2$. For a given $k^\mathrm{th}$ pairing operation, the number of possibilities is reduced to $Q-1-k \times 2$. The number of possibilities for pairing operators are reported in figure \ref{fig:pairing}, according to the number of pairs already constituted (so for the first pair, none has been already constituted before):

\begin{figure}[h!]
\begin{align*}
 1^\mathrm{st} &\longleftrightarrow Q - 1 \\
 2^\mathrm{nd} &\longleftrightarrow Q - 1 - 2 = Q - 3 \\
 3^\mathrm{rd} &\longleftrightarrow Q - 1 - 2 - 2 = Q - 5 \\
 \vdots \\
 k^\mathrm{th} &\longleftrightarrow Q - 1 - 2(k-1) = Q - 2k +1 \\
 \vdots \\
 (M-1)^\mathrm{th} &\longleftrightarrow Q - 1 - 2(M - 1 - 1) = 3 \\
 M^\mathrm{th} &\longleftrightarrow Q - 1 - 2(M-1) = 1
\end{align*}
\caption{Assessment of the number of possibilities one has when pairing two operators, according to the number of operators that have already been paired, i.e., the position of the pairing operation (left side of the figure) in the pairing sequence.}
\label{fig:pairing}
\end{figure}
The number $\mathscr{N}_Q$ of possible products of integrals for a given $Q$-operator is therefore
\begin{equation}
\mathscr{N}_Q = \prod _{k=1} ^{M } (Q - l_k) , \quad l_k = 2k - 1. \label{eq:NQ}
\end{equation}
Since $2k - 1 = Q - 2\left(M-k+1\right) +1$, $\mathscr{N}_Q$ can be rewritten
\begin{equation}
\mathscr{N}_Q = \prod _{k=1}^M (2k-1) = (Q-1)!! \label{eq:Q-1doublef}
\end{equation}
where ``$!!$" denotes a double factorial. This last result has been deduced by considering the fact that the $k^\mathrm{th}$ factor in Eq. \eqref{eq:Q-1doublef} is the $\left(M-k+1\right)^\mathrm{th}$ factor in Eq. \eqref{eq:NQ}. Since $Q = 2M$ is necessarily even, we can finally write
\begin{equation}
\mathscr{N}_Q = (2M-1)!! = \dfrac{(2M)!}{M! \, 2^M} .
\end{equation}
\subsection{Six-operator integral decomposition into a sum of fifteen two-operator integrals}\label{app:3-pair}
\begin{figure}[h!!]
\begin{center}
\includegraphics[scale=0.94]{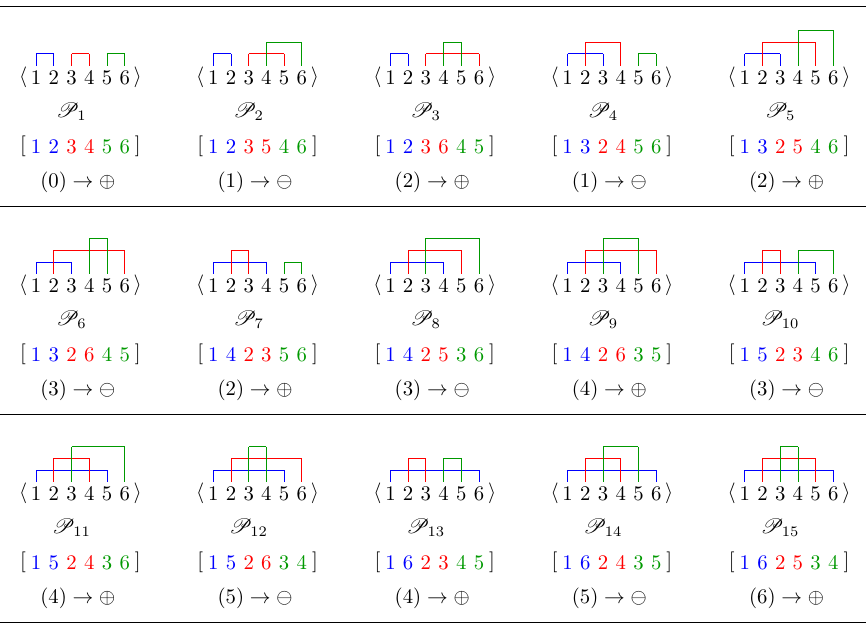}
\end{center}
\caption{Decomposition of a six-operator integral into the sum of fifteen integrals products $\mathscr{P}_{h}$. The number between parentheses is the number $m_h$ of times a pairing line crosses others (the number of times a line of a given color crosses lines with different colors) and the sign on the right of the arrow is the $\displaystyle (-1)^{m_h}$ signature. The reordered labels of the operators are reported between square brackets.}\label{fig:Pd_to_Pr}
\end{figure}

In figure \ref{fig:Pd_to_Pr}, we report the different products of integrals that can be constructed for decomposing a six-operator integral according to Eq. \eqref{eq:initial_I}.
\subsection{Application of the EOM formalism to the calculation of the transition energies}\label{app:EOM}
We start by applying the $\left[ \hat{\mathcal{H}},\hat{\mathcal{T}} \right]$ superoperator to $\ket{\psi _0}$
\begin{equation}
\left.\left.\left[ \hat{\mathcal{H}},\hat{\mathcal{T}} \right] \right| \!\psi _0 \!\right\rangle = \hat{\mathcal{H}}\underbrace{\hat{\mathcal{T}} \ket{\psi _0}}_{\ket{\psi _n}} - \hat{\mathcal{T}} \underbrace{\hat{\mathcal{H}} \ket{\psi _0}}_{\mathscr{E} _0 \ket{\psi _0}} = \hat{\mathcal{H}}\ket{\psi _n} - \mathscr{E}_0 \, \hat{\mathcal{T}} \ket{\psi _0} = \underbrace{\left(\mathscr{E} _n - \mathscr{E} _ 0\right)}_{\omega} \ket{\psi _n} = \omega \, \hat{\mathcal{T}} \ket{\psi _0}.
\end{equation}
This result, together with $\braket{\psi _0 | \psi _n} = \braket{\psi _n| \psi _0} = 0$, can simplify the assessment of the transition energy when replacing $\hat{\mathcal{O}}$ by $\hat{\mathcal{H}}$ in Eq. \eqref{eq:eom-diff}:
\begin{equation}
\left\langle \!\psi _0 \! \left| \left[ \hat{\mathcal{T}}^\dag , \left[ \hat{\mathcal{H}},\hat{\mathcal{T}}\right]\right] \right| \! \psi _0 \! \right\rangle = \left\langle \!\psi _0 \! \left| \left[ \hat{\mathcal{T}}^\dag , \omega \hat{\mathcal{T}} \right] \right| \! \psi _0 \! \right\rangle = \omega \left( \underbrace{\braket{\psi _0 | \hat{\mathcal{T}} ^\dag \hat{\mathcal{T}} | \psi _0}}_{\braket{\psi_0|\psi _0}\braket{\psi_0|\psi_0}} - \underbrace{\braket{\psi _0 | \hat{\mathcal{T}}  \hat{\mathcal{T}}^\dag | \psi _0}}_{\braket{\psi_0|\psi_n}\braket{\psi_n | \psi _0}}\right) = \omega
\end{equation}
If instead we used the full development following Eq. \eqref{eq:eom-diff}, we would obtain
\begin{equation}
\mathscr{D}^{\hat{\mathcal{H}}} _{\mathrm{I}} = \braket{\psi _n | \hat{\mathcal{H}} | \psi _n} = \mathscr{E} _n , \quad \mathscr{D}^{\hat{\mathcal{H}}} _{\mathrm{II}} = \braket{\psi _0 | \hat{\mathcal{H}} | \psi _0} = \mathscr{E} _0
\end{equation}
and
\begin{equation}
\mathscr{D}^{\hat{\mathcal{H}}} _{\mathrm{III}} = \braket{\psi _0 | \hat{\mathcal{H}} | \psi _n}\braket{\psi _n | \psi _0} = 0 = \braket{\psi _0 | \psi _n}\braket{\psi _0 | \hat{\mathcal{H}} | \psi _0}\braket{\psi_n | \psi _0} = \mathscr{D}^{\hat{\mathcal{H}}} _{\mathrm{IV}}
\end{equation}
which finally gives
\begin{equation}
\left\langle \!\psi _0 \! \left| \left[ \hat{\mathcal{T}}^\dag , \left[ \hat{\mathcal{H}},\hat{\mathcal{T}}\right]\right] \right| \! \psi _0 \! \right\rangle = \mathscr{E} _n - \mathscr{E} _0 = \omega .
\end{equation}
\subsection{1--RDM of single-determinant wave functions}\label{app:SD1-RDM}
An $N$--electron Slater determinant
\begin{eqnarray}
\psi (\textbf{s}_1, \dotsc , \textbf{s}_N) = \dfrac{1}{\sqrt{N !}}
\left|
\begin{array}{cccccc}
\varphi _1 (\textbf{s}_1) & \varphi _2 (\textbf{s}_1) & \dots & \varphi _r (\textbf{s}_1) & \dots & \varphi _N (\textbf{s}_1)\\
\varphi _1 (\textbf{s}_2) & \varphi _2 (\textbf{s}_2) & \dots & \varphi _r (\textbf{s}_2) & \dots & \varphi _N (\textbf{s}_2)\\
\vdots & \vdots & & \vdots & &   \vdots \\
\varphi _1 (\textbf{s}_N) & \varphi _2 (\textbf{s}_N) & \dots & \varphi _r (\textbf{s}_N) & \dots & \varphi _N (\textbf{s}_N)\\
\end{array}
\right|  
\end{eqnarray}
can be developed along the $1^\mathrm{st}$ line: We have
\begin{equation}
\psi (\textbf{s}_1, \dotsc , \textbf{s}_N) = \dfrac{1}{\sqrt{N !}} \left\lbrace \sum _{i=1}^N (-1)^{1+i} \mathrm{det}\left( \bm{\psi}^{1i}\right)  \varphi _i (\textbf{s}_1) \right\rbrace  \label{eq:SD-rcol}
\end{equation}
where, if $ \bm{\psi}$ is the matrix from which $\sqrt{N !} \,\psi (\textbf{s}_1, \dotsc , \textbf{s}_N)$ is obtained by taking its determinant, $\bm{\psi}^{kr}$ is the matrix obtained by suppressing the $k^\mathrm{th}$ line and the $r^{\mathrm{th}}$ column of $\bm{\psi}$.  

The 1--RDM kernel is a function written as
\begin{equation}
\gamma (\textbf{s}_1;\textbf{s}_1') =  N   \int \! d\textbf{s}_2 \cdots  \int \! d\textbf{s}_N \; \psi (\textbf{r}_1, \sigma _1 ,..., \textbf{s}_N)  \, \psi  (\textbf{r}_1', \sigma _1', ... ,\textbf{s}_N)\label{eq:kernel}
\end{equation}
so that
\begin{equation}
\sum _{\sigma _1' = \alpha , \beta}\sum _{\sigma _1 = \alpha , \beta} \gamma (\textbf{s}_1;\textbf{s}_1) = n(\textbf{r}_1)
\end{equation}
where $n$ is the one-electron charge density function. It is assumed that, in an $L$-spinorbital basis ($L \geq N$), there exists a 1--RDM, $\bm{\gamma}$, so that the kernel is written as a linear combination of products between spinorbital functions and 1--RDM elements
\begin{equation}
\gamma (\textbf{s}_1;\textbf{s}_1') = \sum _{r=1} ^L \sum _{s=1}^L \varphi _r(\textbf{s}_1)\, (\bm{\gamma}) _{rs} \, \varphi _s  (\textbf{s}_1'). 
\end{equation}
Since it is imposed that, given $T=\left\lbrace1,\dotsc ,L\right\rbrace$,
\begin{equation}
\forall(r,s) \in T^2, \quad \braket{\varphi _r | \varphi _s} = \delta _{rs},
\end{equation}
we have 
\begin{equation}
(\bm{\gamma}) _{r,s} = \int d\textbf{s}_1 \int d\textbf{s}_1' \; \varphi  _r (\textbf{s}_1) \, \varphi _s (\textbf{s}_1') \, \gamma (\textbf{s}_1;\textbf{s}_1') . \label{eq:gammars}
\end{equation}
According to the development of the Slater determinant along the first line given in Eq. \eqref{eq:SD-rcol}, we can write the product of the two $N$--electron one-determinant wave functions in the 1--RDM kernel as
\begin{equation}
\psi (\textbf{r}_1, \sigma _1 ,..., \textbf{s}_N)  \, \psi  (\textbf{r}_1', \sigma _1', ... ,\textbf{s}_N) = (N!)^{-1}\left\lbrace \sum _{{\textcolor{white}{j}}i=1}^N (-1)^{1+i} \mathrm{det}\left( \bm{\psi}^{1i}\right)  \varphi _i (\textbf{s}_1) \right\rbrace \left\lbrace \sum _{j=1}^N (-1)^{1+j} \mathrm{det}\left( \bm{\psi}^{1j \dag}\right)  \varphi _j (\textbf{s}_1') \right\rbrace
\end{equation}
with $\textbf{s}_1 = (\textbf{r}_1,\sigma _1)$ and $\textbf{s}_1' = (\textbf{r}_1',\sigma _1')$. This can be rewritten
\begin{equation} 
\psi (\textbf{s}_1 ,..., \textbf{s}_N)  \, \psi  (\textbf{s}_1',  ... ,\textbf{s}_N) = (N!)^{-1}\left\lbrace \sum _{i=1}^N\sum _{j=1}^N (-1)^{i+j} \mathrm{det}\left( \bm{\psi}^{1i}\right)  \mathrm{det}\left( \bm{\psi}^{1j \dag}\right) \varphi _i (\textbf{s}_1)  \varphi _j (\textbf{s}_1') \right\rbrace 
\end{equation}
where $ \mathrm{det}\left( \bm{\psi}^{1i}\right) $ and $\mathrm{det}\left( \bm{\psi}^{1j \dag}\right)$ correspond to the determinants that can be used for writing two arbitrary $(N-1)$-electron wave functions
\begin{equation}
\psi ^{1i} (\textbf{s}_2, ..., \textbf{s}_N) = \dfrac{1}{\sqrt{(N-1)!}} \, \mathrm{det}\left( \bm{\psi}^{1i}\right)
\end{equation}
and
\begin{equation}
\psi ^{1j} (\textbf{s}_2, ..., \textbf{s}_N) = \dfrac{1}{\sqrt{(N-1)!}} \, \mathrm{det}\left( \bm{\psi}^{1j \dag}\right).
\end{equation}
According to the fact that the integral over all the space of the product between one wave function and its complex conjugate is equal to one, and that two one-determinant wave functions differing by one spinorbital have a zero spatial overlap (see appendix \ref{app:SDoverlap}), 
\begin{equation}
\int \! d\textbf{s}_2 \cdots \int \! d\textbf{s}_N \, \psi ^{1i} (\textbf{s}_2, ..., \textbf{s}_N) \psi ^{1j} (\textbf{s}_2, ..., \textbf{s}_N) = \delta _{ij} \Longleftrightarrow \int \! d\textbf{s}_2 \cdots \int \! d\textbf{s}_N \, \mathrm{det}\left( \bm{\psi}^{1i}\right) \mathrm{det}\left( \bm{\psi}^{1j \dag}\right) = \delta _{ij}\,(N-1)!
\end{equation}
Therefore, integrating over ($N-1$) spin-spatial coordinates gives
\begin{equation}
\int \! d\textbf{s}_2 \cdots \int \! d\textbf{s}_N  \; \psi (\textbf{s}_1 ,..., \textbf{s}_N)  \, \psi  (\textbf{s}_1', ... ,\textbf{s}_N) = (N!)^{-1} \sum _{i=1}^N\sum _{j=1}^N (-1)^{i+j} \varphi _i (\textbf{s}_1) \varphi _j (\textbf{s}_1')  \, \delta _{ij} \, (N-1)!
\end{equation}
which can be simplified:
\begin{equation}
  N  \int \! d\textbf{s}_2 \cdots \int \! d\textbf{s}_N  \; \psi (\textbf{s}_1 ,..., \textbf{s}_N)  \, \psi  (\textbf{s}_1', ... ,\textbf{s}_N) = \sum _{i=1}^N \varphi _i(\textbf{s}_1)\varphi _{i} (\textbf{s}_1')
\end{equation}
where the left hand-side is nothing but the 1--RDM kernel $\gamma (\textbf{s}_1;\textbf{s}_1')$ written as in Eq. \eqref{eq:kernel}. Multiplying now by $\varphi  _r (\textbf{s}_1)$ and $\varphi  _s (\textbf{s}_1')$ and integrating as in Eq. \eqref{eq:gammars} returns
\begin{equation}
(\bm{\gamma}) _{r,s} = \int d\textbf{s}_1 \int d\textbf{s}_1' \; \varphi  _r (\textbf{s}_1) \, \varphi _s (\textbf{s}_1') \, \sum _{i=1}^N \varphi _i(\textbf{s}_1)\varphi _{i} (\textbf{s}_1') = \sum _{i=1}^N\delta _{ri} \delta_{si} =  \delta _{rs}n_r n_s = \delta _{rs}n_s = \braket{\psi | \hat{r}^\dag\hat{s} | \psi} ,
\end{equation}
i.e.,
\begin{equation}
\bm{\gamma} = I_o \oplus 0_v .
\end{equation}
\subsection{Overlap between two Slater determinants differing by one spinorbital}\label{app:SDoverlap}
We explicitly define the $\bm{\psi}$ matrix elements as
\begin{equation}
(\bm{\psi})_{i,j} := \varphi _j (\textbf{s}_i)
\end{equation}
such that
\begin{equation}
\psi (\textbf{s}_1, \dotsc , \textbf{s}_N) = \dfrac{1}{\sqrt{N!}} \left| \bm{\psi} \right|
\end{equation}
and the $\bm{\psi}_j ^a$ matrix such that, for every $i \, \in \left\lbrace 1, ..., N\right\rbrace$, $\varphi _j (\textbf{s}_i)$ in $\bm{\psi}$ has been substituted by $\varphi _a (\textbf{s}_i)$, with $1 \leq j \leq N$ and $N < a \leq L$.
Then, according to 
\begin{equation}
 \mathrm{det}(\textbf{A}) \mathrm{det}(\textbf{B}) = \mathrm{det}(\textbf{AB}),
\end{equation}
we have
 \begin{equation}
 \psi (\textbf{s}_1, \dotsc , \textbf{s}_N)\psi _j^{a}(\textbf{s}_1, \dotsc , \textbf{s}_N)  = (N!)^{-1}\mathrm{det}\left( \bm{\psi}_{\textcolor{white}{j}}^{\textcolor{white}{a}}\!\!\right) \mathrm{det}\left( \bm{\psi}_j ^{a}\right)  .
\end{equation}
We therefore have
\begin{equation}
\left(\bm{\psi}\bm{\psi}_j^{a}\right)_{ij} = \sum _{k=1}^N  {\left(\bm{\psi}\right)_{ik}}{\left(\bm{\psi}_j^{a}\right)_{kj}} = \sum _{k=1}^N\varphi _k (\textbf{s}_i) \varphi _a(\textbf{s}_k),
\end{equation}
which leads to
\begin{equation}
\mathrm{det}\left(\bm{\psi}\bm{\psi}_j^{a} \right)  = \sum _{i=1}^N (-1)^{i+j} \left( \bm{\psi} \bm{\psi}_j^{a}\right)_{ij} \mathpzc{D}(i|j)= \sum _{i=1}^N \sum _{k=1}^N (-1)^{i+j} \varphi _k (\textbf{s}_i) \varphi _a(\textbf{s}_k)\mathpzc{D}(i|j).
\end{equation}
where the sub-determinant $\mathpzc{D}(i|j)$ is obtained by subtracting the $i^\mathrm{th}$ line and the $j^\mathrm{th}$ column from $\bm{\psi}\bm{\psi}_j^{a}$.

Integrating the wave function product reads
\begin{align}
\nonumber S_j^a &=  \int \! d\textbf{s}_1 \cdots \int \! d\textbf{s}_N \, \psi (\textbf{s}_1, \dotsc , \textbf{s}_N)\psi _j^{a}(\textbf{s}_1, \dotsc , \textbf{s}_N) \\  &= \vspace*{-1cm}(N!)^{-1} \int \! d\textbf{s}_1 \cdots \int \! d\textbf{s}_N\sum _{i=1}^N (-1)^{i+j} \sum _{k=1}^N \varphi _k (\textbf{s}_i) \varphi _a(\textbf{s}_k)  \mathpzc{D}(i|j) .
\end{align}
For every occurrence of $i = k$, we have 
\begin{align}
 (N!)^{-1}  (-1)^{i+j} \int \! d\textbf{s}_1 \cdots \int \! d\textbf{s}_N \, \mathpzc{D}(i|j)  \varphi _i (\textbf{s}_i) \varphi _a (\textbf{s}_i)  \propto \int d\textbf{s}_i \, \varphi _i (\textbf{s}_i) \varphi _a (\textbf{s}_i)  = \delta _{ia} = 0.
\end{align}
For other occurrences ($i \neq k$), we  must develop $\mathpzc{D}(i|j)$. Since $\varphi _a$ will always be multiplied by an occupied spinorbital before integration, each term of the overlap integral $S_j^a$ will be vanishing.
\subsection{The case of ADC excited states}\label{app:ADC}
Using the intermediate state representation in ADC leads to the following excited-state ansatz:
\begin{equation}
\ket{\psi _n^\mathrm{ADC}} = \sum _{J}\, (\textbf{X}^\mathrm{ADC})_{Jn}\ket{\tilde{\Psi} _J}
\end{equation}
with the $\ket{\tilde{\Psi}_J}$ being the so-called ``intermediate states", forming the space of intermediate states within which the ADC ansatz is defined, with the property that
\begin{equation}\label{eq:ISRprop}
\braket{\tilde{\Psi}_I|\tilde{\Psi}_J} = \delta_{IJ}, \quad \braket{\tilde{\Psi}_J|\Psi_0} = \braket{\Psi _0|\tilde{\Psi}_J} = 0.
\end{equation}
where $\ket{\Psi _0}$ is the correlated ground state reference used in ADC. On the other hand, equation \eqref{eq:Tzetapostulate} could be rewritten
\begin{equation}
 \ket{\psi _n^\zeta} = \sum _{i=1}^N \sum _{a=N+1}^L (\tilde{\textbf{X}}_\zeta)_{ia} \ket{\psi _i ^a} 
\end{equation}
with the singly-excited Slater determinants $\ket{\psi _i ^a}$ forming another space with the same properties as those reported for the intermediate states in equation \eqref{eq:ISRprop}, i.e.,
\begin{equation}
\braket{{\psi}_i^a|{\psi}_j^b} = \delta_{ij}\delta_{ab}, \quad \braket{{\psi}_i^a|\psi_0} = \braket{\psi _0|{\psi}_i^a} = 0.
\end{equation}
Since the structure of both ansätze is identical, and the two related spaces have the same vectorial properties, we can conclude that an electronic transition within the ADC framework will have a matrix representation (transition and difference density matrix) in the space of intermediate states with a structure identical to the matrix representation of an electronic transition using the $ \ket{\psi _n^\zeta}$ ansatz in the space of singly-excited Slater determinants.

\bibliographystyle{ieeetr}

\begin{thebibliography}{100}

\bibitem{dieter_optogenetics_2019}
A.~Dieter, C.J.~Duque-Afonso, V.~Rankovic, M.~Jeschke and T.~Moser, ``Near physiological spectral selectivity of cochlear optogenetics,"{\em Nat. Commun.}, 10, art.~1962, 2019.

\bibitem{youngsik_therapy_2019}
L.~Youngsik, K.~Dae-Hyeong, ``Wireless metronomic photodynamic therapy," {\em Nat. Biomed. Eng.}, vol.~3, pp.~5--6.

\bibitem{editorial_perovskite_2019}
{\em Editorial: }``A decade of perovskite photovoltaics," {\em Nat. Energy}, vol.~4, p.~1, 2019.

\bibitem{dreuw_single-reference_2005}
A.~Dreuw and M.~Head-Gordon, ``Single-{Reference} ab {Initio} {Methods} for the
  {Calculation} of {Excited} {States} of {Large} {Molecules},'' {\em Chemical
  Reviews}, vol.~105, pp.~4009--4037, Nov. 2005.

\bibitem{plasser_new_2014}
F.~Plasser, M.~Wormit, and A.~Dreuw, ``New tools for the systematic analysis
  and visualization of electronic excitations. i. formalism,'' {\em J. Chem.
  Phys.}, vol.~141, p.~024106, July 2014.

\bibitem{savarese_metrics_2017}
M.~Savarese, C.~A. Guido, E.~Brémond, I.~Ciofini, and C.~Adamo, ``Metrics for
  {Molecular} {Electronic} {Excitations}: {A} {Comparison} between {Orbital}-
  and {Density}-{Based} {Descriptors},'' {\em The Journal of Physical Chemistry
  A}, vol.~121, pp.~7543--7549, Oct. 2017.

\bibitem{luzanov_electron_2010}
A.~V. Luzanov and O.~A. Zhikol, ``Electron invariants and excited state
  structural analysis for electronic transitions within cis, rpa, and tddft
  models,'' {\em Int. J. Quantum Chem.}, vol.~110, no.~4, pp.~902--924, 2010.

\bibitem{luzanov_interpretation_1980}
A.~V. Luzanov and V.~F. Pedash, ``Interpretation of excited states using
  charge-transfer numbers,'' {\em Theor. Exp. Chem.}, vol.~15, pp.~338--341,
  July 1980.

\bibitem{furche_density_2001}
F.~Furche, ``On the density matrix based approach to time-dependent density
  functional response theory,'' {\em The Journal of Chemical Physics},
  vol.~114, pp.~5982--5992, Mar. 2001.

\bibitem{tretiak_density_2002}
S.~Tretiak and S.~Mukamel, ``Density matrix analysis and simulation of
  electronic excitations in conjugated and aggregated molecules,'' {\em Chem.
  Rev.}, vol.~102, pp.~3171--3212, Sept. 2002.

\bibitem{martin_natural_2003}
R.~L. Martin, ``Natural transition orbitals,'' {\em J. Chem. Phys.}, vol.~118,
  no.~11, pp.~4775--4777, 2003.

\bibitem{batista_et_martin_natural_2004}
E.~R. Batista and R.~L. Martin, ``Natural transition orbitals,'' {\em
  Encyclopedia of Computational Chemistry}, P. Ragué Schleyer, N. L. Allinger,
  T. Clark, J. Gasteiger, P. A. Kollman, H. F. Schaefer, P. R. Schreiner, W.
  Thiel, W. L. Jorgensen and R. C. Glen Eds., American Cancer Society, 2004.

\bibitem{tretiak_exciton_2005}
S.~Tretiak, K.~Igumenshchev, and V.~Chernyak, ``Exciton sizes of conducting
  polymers predicted by time-dependent density functional theory,'' {\em
  Physical review B}, vol.~71, no.~3, p.~033201, 2005.

\bibitem{mayer_using_2007}
I.~Mayer, ``Using singular value decomposition for a compact presentation and
  improved interpretation of the cis wave functions,'' {\em Chem. Phys. Lett.},
  vol.~437, pp.~284--286, Apr. 2007.

\bibitem{surjan_natural_2007}
P.~R. Surján, ``Natural orbitals in cis and singular-value decomposition,''
  {\em Chem. Phys. Lett.}, vol.~439, pp.~393--394, May 2007.

\bibitem{wu_exciton_2008}
C.~Wu, S.~V. Malinin, S.~Tretiak, and V.~Y. Chernyak, ``Exciton scattering
  approach for branched conjugated molecules and complexes. iii.
  applications,'' {\em J. Chem. Phys.}, vol.~129, no.~17, p.~174113, 2008.

\bibitem{li_time-dependent_2011}
Y.~Li and C.~A. Ullrich, ``Time-dependent transition density matrix,'' {\em
  Chemical Physics}, vol.~391, pp.~157--163, Nov. 2011.

\bibitem{plasser_analysis_2012}
F.~Plasser and H.~Lischka, ``Analysis of excitonic and charge transfer
  interactions from quantum chemical calculations,'' {\em J. Chem. Theory
  Comput.}, vol.~8, pp.~2777--2789, Aug. 2012.

\bibitem{pluhar_visualizing_2018}
E.~A. Pluhar and C.~A. Ullrich, ``Visualizing electronic excitations with the
  particle-hole map: orbital localization and metric space analysis,'' {\em
  Eur. Phys. J. B}, vol.~91, p.~137, July 2018.

\bibitem{plasser_newbis_2014}
F.~Plasser, S.~A. Bäppler, M.~Wormit, and A.~Dreuw, ``New tools for the
  systematic analysis and visualization of electronic excitations. ii.
  applications,'' {\em J. Chem. Phys.}, vol.~141, p.~024107, July 2014.

\bibitem{bappler_exciton_2014}
S.~A. Bäppler, F.~Plasser, M.~Wormit, and A.~Dreuw, ``Exciton analysis of
  many-body wave functions: Bridging the gap between the quasiparticle and
  molecular orbital pictures,'' {\em Phys. Rev. A}, vol.~90, p.~052521, Nov.
  2014.

\bibitem{mewes2015communication}
S.~A. Mewes, F.~Plasser, and A.~Dreuw, ``Communication: Exciton analysis in
  time-dependent density functional theory: How functionals shape excited-state
  characters,'' {\em J. Chem. Phys.}, vol.~143, no.~17, p.~171101, 2015.

\bibitem{li_particlehole_2015}
Y.~Li and C.~A. Ullrich, ``The {Particle}–{Hole} {Map}: {A} {Computational}
  {Tool} {To} {Visualize} {Electronic} {Excitations},'' {\em Journal of
  Chemical Theory and Computation}, vol.~11, pp.~5838--5852, Dec. 2015.
  
\bibitem{li_particlehole_2016}
Y.~Li and C.~A. Ullrich, ``The particle-hole map: Formal derivation and numerical implementation," {\em J. Chem. Phys.}, vol.~145, p.~164107,~2016.

\bibitem{etienne2015transition}
T.~Etienne, ``Transition matrices and orbitals from reduced density matrix
  theory,'' {\em J. Chem. Phys.}, vol.~142, no.~24, p.~244103, 2015.

\bibitem{plasser2015statistical}
F.~Plasser, B.~Thomitzni, S.~A. B{\"a}ppler, J.~Wenzel, D.~R. Rehn, M.~Wormit,
  and A.~Dreuw, ``Statistical analysis of electronic excitation processes:
  Spatial location, compactness, charge transfer, and electron-hole
  correlation,'' {\em J. Comp. Chem.}, vol.~36, no.~21, pp.~1609--1620, 2015.

\bibitem{poidevin_truncated_2016}
C.~Poidevin, C.~Lepetit, N.~Ben~Amor, and R.~Chauvin, ``Truncated {Transition}
  {Densities} for {Analysis} of ({Nonlinear}) {Optical} {Properties} of
  carbo-{Chromophores},'' {\em Journal of Chemical Theory and Computation},
  vol.~12, pp.~3727--3740, Aug. 2016.

\bibitem{wenzel_physical_2016}
J.~Wenzel and A.~Dreuw, ``Physical {Properties}, {Exciton} {Analysis}, and
  {Visualization} of {Core}-{Excited} {States}: {An} {Intermediate} {State}
  {Representation} {Approach},'' {\em J. Chem. Theory Comput.}, vol.~12,
  pp.~1314--1330, Mar. 2016.

\bibitem{plasser2016entanglement}
F.~Plasser, ``Entanglement entropy of electronic excitations,'' {\em J. Chem.
  Phys.}, vol.~144, no.~19, p.~194107, 2016.

\bibitem{plasser_detailed_2017}
F.~Plasser, S.~A. Mewes, A.~Dreuw, and L.~González, ``Detailed {Wave}
  {Function} {Analysis} for {Multireference} {Methods}: {Implementation} in the
  {Molcas} {Program} {Package} and {Applications} to {Tetracene},'' {\em
  Journal of Chemical Theory and Computation}, vol.~13, pp.~5343--5353, Nov.
  2017.

\bibitem{mai_quantitative_2018}
S.~Mai, F.~Plasser, J.~Dorn, M.~Fumanal, C.~Daniel, and L.~Gonzalez,
  ``Quantitative wave function analysis for excited states of transition metal
  complexes,'' {\em Coord. Chem. Rev.}, vol.~361, pp.~74--97, Apr. 2018.
\newblock arXiv: 1711.10707.

\bibitem{mewes_benchmarking_2018}
S.~A. Mewes, F.~Plasser, A.~Krylov, and A.~Dreuw, ``Benchmarking
  {Excited}-{State} {Calculations} {Using} {Exciton} {Properties},'' {\em J.
  Chem. Theory Comput.}, vol.~14, pp.~710--725, Feb. 2018.

\bibitem{etienne_theoretical_2018}
T.~Etienne, ``Theoretical {Insights} into the {Topology} of {Molecular}
  {Excitons} from {Single}-{Reference} {Excited} {States} {Calculation}
  {Methods},'' {\em Excitons}, Sergei L. Pyshkin Ed., InTech, 2018, doi
  10.5772/intechopen.70688.

\bibitem{skomorowski_real_2018}
W.~Skomorowski and A.~I. Krylov, ``Real and {Imaginary} {Excitons}: {Making}
  {Sense} of {Resonance} {Wave} {Functions} by {Using} {Reduced} {State} and
  {Transition} {Density} {Matrices},'' {\em The Journal of Physical Chemistry
  Letters}, vol.~9, pp.~4101--4108, July 2018.

\bibitem{park_low_2018}
Y.~C. Park, A.~Perera, and R.~J. Bartlett, ``Low scaling {EOM}-{CCSD} and
  {EOM}-{MBPT}(2) method with natural transition orbitals,'' {\em The Journal
  of Chemical Physics}, vol.~149, p.~184103, Nov. 2018.

\bibitem{HeadGordonJPhysChem1995}
M.~Head-Gordon, A.~M. Grana, D.~Maurice, and C.~A. White, ``Analysis of
  electronic transitions as the difference of electron attachment and
  detachment densities,'' {\em J. Phys. Chem.}, vol.~99, p.~14261, 1995.

\bibitem{ronca_charge-displacement_2014}
E.~Ronca, M.~Pastore, L.~Belpassi, F.~D. Angelis, C.~Angeli, R.~Cimiraglia, and
  F.~Tarantelli, ``Charge-displacement analysis for excited states,'' {\em J.
  Chem. Phys.}, vol.~140, p.~054110, Feb. 2014.

\bibitem{ronca_density_2014}
E.~Ronca, C.~Angeli, L.~Belpassi, F.~De~Angelis, F.~Tarantelli, and M.~Pastore,
  ``Density relaxation in time-dependent density functional theory: Combining
  relaxed density natural orbitals and multireference perturbation theories for
  an improved description of excited states,'' {\em J. Chem. Theory Comput.},
  vol.~10, no.~9, pp.~4014--4024, 2014.

\bibitem{etienne_new_2014}
T.~Etienne, X.~Assfeld, and A.~Monari, ``New insight into the topology of
  excited states through detachment/attachment density matrices-based centroids
  of charge,'' {\em J. Chem. Theory Comput.}, vol.~10, pp.~3906--3914, Sept.
  2014.

\bibitem{pastore2017unveiling}
M.~Pastore, X.~Assfeld, E.~Mosconi, A.~Monari, and T.~Etienne, ``Unveiling the
  nature of post-linear response z-vector method for time-dependent density
  functional theory,'' {\em J. Chem. Phys.}, vol.~147, no.~2, p.~024108, 2017.

\bibitem{peach_excitation_2008}
M.~J.~G. Peach, P.~Benfield, T.~Helgaker, and D.~J. Tozer, ``Excitation
  energies in density functional theory: an evaluation and a diagnostic test,''
  {\em J. Chem. Phys.}, vol.~128, p.~044118, Jan. 2008.

\bibitem{le_bahers_qualitative_2011}
T.~Le~Bahers, C.~Adamo, and I.~Ciofini, ``A qualitative index of spatial extent
  in charge-transfer excitations,'' {\em J. Chem. Theory Comput.}, vol.~7,
  pp.~2498--2506, Aug. 2011.

\bibitem{garcia_evaluating_2013}
G.~García, C.~Adamo, and I.~Ciofini, ``Evaluating push–pull dye efficiency
  using td-dft and charge transfer indices,'' {\em Phys. Chem. Chem. Phys.},
  Oct. 2013.

\bibitem{GuidoJChemTheoryComput2013}
C.~A. Guido, P.~Cortona, B.~Mennucci, and C.~Adamo, ``On the metric of charge
  transfer molecular excitations: A simple chemical descriptor,'' {\em J. Chem.
  Theory Comput.}, vol.~9, no.~7, pp.~3118--3126, 2013.

\bibitem{etienne_toward_2014}
T.~Etienne, X.~Assfeld, and A.~Monari, ``Toward a quantitative assessment of
  electronic transitions' charge-transfer character,'' {\em J. Chem. Theory
  Comput.}, vol.~10, pp.~3896--3905, Sept. 2014.

\bibitem{guido_effective_2014}
C.~A. Guido, P.~Cortona, and C.~Adamo, ``Effective electron displacements: A
  tool for time-dependent density functional theory computational
  spectroscopy,'' {\em J. Chem. Phys.}, vol.~140, p.~104101, Mar. 2014.

\bibitem{etienne_probing_2015}
T.~Etienne, ``Probing the {Locality} of {Excited} {States} with {Linear}
  {Algebra},'' {\em Journal of Chemical Theory and Computation}, vol.~11,
  pp.~1692--1699, Apr. 2015.

\bibitem{barca_excitation_2018}
G.~M.~J. Barca, A.~T.~B. Gilbert, and P.~M.~W. Gill, ``Excitation {Number}:
  {Characterizing} {Multiply} {Excited} {States},'' {\em Journal of Chemical
  Theory and Computation}, vol.~14, pp.~9--13, Jan. 2018.

\bibitem{campetella_quantifying_2019}
M.~Campetella, A.~Perfetto, and I.~Ciofini, ``Quantifying partial hole-particle
  distance at the excited state: {A} revised version of the {DCT} index,'' {\em
  Chemical Physics Letters}, vol.~714, pp.~81--86, Jan. 2019.

\bibitem{cioslowski_many-electron_2000}
J.~Cioslowski, ed., {\em Many-{Electron} {Densities} and {Reduced} {Density}
  {Matrices}}.
\newblock Mathematical and {Computational} {Chemistry}, Springer US, 2000.

\bibitem{coleman_structure_1963}
A.~J. Coleman, ``Structure of {Fermion} {Density} {Matrices},'' {\em Reviews of
  Modern Physics}, vol.~35, pp.~668--686, July 1963.

\bibitem{davidson_reduced_1976}
E.~Davidson, {\em Reduced {Density} {Matrices} in {Quantum} {Chemistry}},
  vol.~6 of {\em Theoretical {Chemistry}}.
\newblock Elsevier, 1976.

\bibitem{lowdin_quantum_1955}
P.-O. Löwdin, ``Quantum {Theory} of {Many}-{Particle} {Systems}. {I}.
  {Physical} {Interpretations} by {Means} of {Density} {Matrices}, {Natural}
  {Spin}-{Orbitals}, and {Convergence} {Problems} in the {Method} of
  {Configurational} {Interaction},'' {\em Physical Review}, vol.~97,
  pp.~1474--1489, Mar. 1955.

\bibitem{mcweeny_recent_1960}
R.~McWeeny, ``Some {Recent} {Advances} in {Density} {Matrix} {Theory},'' {\em
  Reviews of Modern Physics}, vol.~32, pp.~335--369, Apr. 1960.

\bibitem{mcweeny_methods_1992}
R.~McWeeny, {\em Methods of molecular quantum mechanics}.
\newblock Theoretical chemistry, London: Academic Press, 2nd ed~ed., 1992.

\bibitem{mcweeny_density_1959}
R.~McWeeny, ``The density matrix in many-electron quantum mechanics {I}.
  {Generalized} product functions. {Factorization} and physical interpretation
  of the density matrices,'' {\em Proc. R. Soc. Lond. A}, vol.~253,
  pp.~242--259, Nov. 1959.

\bibitem{mcweeny_density_1961}
R.~McWeeny and Y.~Mizuno, ``The density matrix in many-electron quantum
  mechanics {II}. {Separation} of space and spin variables; spin coupling
  problems,'' {\em Proc. R. Soc. Lond. A}, vol.~259, pp.~554--577, Jan. 1961.

\bibitem{wick_evaluation_1950}
G.~C. Wick, ``The {Evaluation} of the {Collision} {Matrix},'' {\em Physical
  Review}, vol.~80, pp.~268--272, Oct. 1950.

\bibitem{gross_many-particle_1991}
E.~K.~U. Gross, E.~Runge, and O.~Heinonen, {\em Many-{Particle} {Theory}.}
\newblock Bristol: Adam Hilger, 1991.
\newblock OCLC: 471769116.

\bibitem{kutzelnigg_normal_1997}
W.~Kutzelnigg and D.~Mukherjee, ``Normal order and extended {Wick} theorem for
  a multiconfiguration reference wave function,'' {\em The Journal of Chemical
  Physics}, vol.~107, pp.~432--449, July 1997.

\bibitem{surjan_second_1989}
P.~R. Surjan, {\em Second {Quantized} {Approach} to {Quantum} {Chemistry}: {An}
  {Elementary} {Introduction}}.
\newblock Berlin Heidelberg: Springer-Verlag, 1989.

\bibitem{szabo1996szabo}
A.~Szabo and N.S.~Ostlund, {\em Modern {Quantum} {Chemistry}: {Introduction} to {Advanced} {Electronic} {Structure} {Theory}}.
\newblock Dover Publications: New York, 1996.

\bibitem{lindgren2012atomic}
I.~Lindgren and J.~Morrison, {\em Atomic many-body theory}.
\newblock Springer Science \& Business Media, 2012.

\bibitem{bogolyubov1980introduction}
N.N.~Bogoliubov and D.V.~Shirkov, {\em Introduction to the theory of quantized fields}.
\newblock Wiley-Interscience Publishers, 1959.

\bibitem{shavitt_many-body_2009}
I.~Shavitt and R.~J. Bartlett, {\em Many-{Body} {Methods} in {Chemistry} and
  {Physics}: {MBPT} and {Coupled}-{Cluster} {Theory}}.
\newblock Cambridge {Molecular} {Science}, Cambridge University Press, Aug.
  2009.

\bibitem{wilson_methods_1992}
S.~Wilson and G.~H.~F. Diercksen, eds., {\em Methods in {Computational}
  {Molecular} {Physics}}.
\newblock Nato {Science} {Series} {B}:, Springer US, 1992.

\bibitem{rowe_equations--motion_1968}
D.~J. Rowe, ``Equations-of-{Motion} {Method} and the {Extended} {Shell}
  {Model},'' {\em Reviews of Modern Physics}, vol.~40, pp.~153--166, Jan. 1968.

\bibitem{yeager_equations_1975}
D.~L. Yeager and V.~McKoy, ``An equations of motion approach for open shell
  systems,'' {\em The Journal of Chemical Physics}, vol.~63, pp.~4861--4869,
  Dec. 1975.

\bibitem{mccurdy_equations_1977}
C.~W. McCurdy, T.~N. Rescigno, D.~L. Yeager, and V.~McKoy, ``The {Equations} of
  {Motion} {Method}: {An} {Approach} to the {Dynamical} {Properties} of {Atoms}
  and {Molecules},'' in {\em Methods of {Electronic} {Structure} {Theory}},
  Modern {Theoretical} {Chemistry}, pp.~339--386, Springer, Boston, MA, 1977.

\bibitem{lynch_excited_1982}
D.~Lynch, M.~F. Herman, and D.~L. Yeager, ``Excited state properties from the
  equations of motion method. {Application} of the {MCTDHF}-{MCRPA} to the
  dipole moments and oscillator strengths of the low-lying valence states of
  {CO},'' {\em Chemical Physics}, vol.~64, pp.~69--81, Jan. 1982.

\bibitem{ehrenreich_self-consistent_1959}
H.~Ehrenreich and M.~H. Cohen, ``Self-{Consistent} {Field} {Approach} to the
  {Many}-{Electron} {Problem},'' {\em Phys. Rev.}, vol.~115, pp.~786--790, Aug.
  1959.

\bibitem{mckoy_equations_1980}
V.~McKoy, ``The {Equations} of {Motion} {Method}: {An} {Approach} to the
  {Dynamical} {Properties} of {Atoms} and {Molecules},'' {\em Phys. Scr.},
  vol.~21, no.~3-4, p.~238, 1980.

\bibitem{joergensen_second_1981}
P.~Joergensen, {\em Second {Quantization}-{Based} {Methods} in {Quantum}
  {Chemistry}}.
\newblock Elsevier, 1981.

\bibitem{schirmer_new_1996}
J.~Schirmer and F.~Mertins, ``A new approach to the random phase
  approximation,'' {\em Journal of Physics B: Atomic, Molecular and Optical
  Physics}, vol.~29, no.~16, p.~3559, 1996.

\bibitem{furche_developing_2008}
F.~Furche, ``Developing the random phase approximation into a practical
  post-{Kohn}–{Sham} correlation model,'' {\em J. Chem. Phys.}, vol.~129,
  p.~114105, Sept. 2008.

\bibitem{toulouse_range-separated_2010}
J.~Toulouse, W.~Zhu, J.~G. Ángyán, and A.~Savin, ``Range-separated
  density-functional theory with the random-phase approximation: {Detailed}
  formalism and illustrative applications,'' {\em Phys. Rev. A}, vol.~82,
  p.~032502, Sept. 2010.

\bibitem{sauer_molecular_2011}
S.~P.~A. Sauer, {\em Molecular {Electromagnetism}: {A} {Computational}
  {Chemistry} {Approach}}.
\newblock Oxford University Press, Aug. 2011.

\bibitem{chatterjee_excitation_2012}
K.~Chatterjee and K.~Pernal, ``Excitation energies from extended random phase
  approximation employed with approximate one- and two-electron reduced density
  matrices,'' {\em The Journal of Chemical Physics}, vol.~137, p.~204109, Nov.
  2012.

\bibitem{pernal_intergeminal_2014}
K.~Pernal, ``Intergeminal {Correction} to the {Antisymmetrized} {Product} of
  {Strongly} {Orthogonal} {Geminals} {Derived} from the {Extended} {Random}
  {Phase} {Approximation},'' {\em Journal of Chemical Theory and Computation},
  vol.~10, pp.~4332--4341, Oct. 2014.

\bibitem{pastorczak_correlation_2018}
E.~Pastorczak and K.~Pernal, ``Correlation {Energy} from the {Adiabatic}
  {Connection} {Formalism} for {Complete} {Active} {Space} {Wave}
  {Functions},'' {\em Journal of Chemical Theory and Computation}, May 2018.

\bibitem{pernal_correlation_2018}
K.~Pernal, ``Correlation energy from random phase approximations: {A} reduced
  density matrices perspective,'' {\em International Journal of Quantum
  Chemistry}, vol.~118, p.~e25462, July 2018.

\bibitem{pernal_electron_2018}
K.~Pernal, ``Electron {Correlation} from the {Adiabatic} {Connection} for
  {Multireference} {Wave} {Functions},'' {\em Physical Review Letters},
  vol.~120, p.~013001, Jan. 2018.

\bibitem{mclachlan_time-dependent_1964}
A.~D. McLachlan and M.~A. Ball, ``Time-{Dependent} {Hartree}-{Fock} {Theory}
  for {Molecules},'' {\em Reviews of Modern Physics}, vol.~36, pp.~844--855,
  July 1964.

\bibitem{dalgaard_timedependent_1980}
E.~Dalgaard, ``Time-dependent multiconfigurational {Hartree}–{Fock} theory,''
  {\em The Journal of Chemical Physics}, vol.~72, pp.~816--823, Jan. 1980.

\bibitem{mcweeny_time-dependent_1983}
R.~Mcweeny, ``Time-dependent hartree-fock theory and its multiconfiguration
  generalization,'' {\em Journal of Molecular Structure: THEOCHEM}, vol.~93,
  pp.~1--14, May 1983.

\bibitem{yeager_generalizations_1984}
D.~L. Yeager, J.~Olsen, and P.~Jørgensen, ``Generalizations of the
  multiconfigurational time-dependent {Hartree}–{Fock} approach,'' {\em
  Faraday Symposia of the Chemical Society}, vol.~19, pp.~85--95, Jan. 1984.

\bibitem{hirata_configuration_1999}
S.~Hirata, M.~Head-Gordon, and R.~J. Bartlett, ``Configuration interaction
  singles, time-dependent {Hartree}–{Fock}, and time-dependent density
  functional theory for the electronic excited states of extended systems,''
  {\em The Journal of Chemical Physics}, vol.~111, pp.~10774--10786, Dec. 1999.

\bibitem{crespo-otero_recent_2018}
R.~Crespo-Otero and M.~Barbatti, ``Recent {Advances} and {Perspectives} on
  {Nonadiabatic} {Mixed} {Quantum}–{Classical} {Dynamics},'' {\em Chemical
  Reviews}, vol.~118, pp.~7026--7068, Aug. 2018.

\bibitem{casida_time-dependent_1995}
M.~E. Casida, ``Time-dependent density functional response theory for
  molecules,'' in {\em Recent {Advances} in {Density} {Functional} {Methods}},
  vol.~Volume 1 of {\em Recent {Advances} in {Computational} {Chemistry}},
  pp.~155--192, World Scientific, Nov. 1995.

\bibitem{jamorski_dynamic_1996}
C.~Jamorski, M.~E. Casida, and D.~R. Salahub, ``Dynamic polarizabilities and
  excitation spectra from a molecular implementation of time-dependent
  density-functional response theory: {N}2 as a case study,'' {\em The Journal
  of Chemical Physics}, vol.~104, pp.~5134--5147, Apr. 1996.

\bibitem{casida_time-dependent_1996}
M.~E. Casida, ``Time-{Dependent} {Density} {Functional} {Response} {Theory} of
  {Molecular} {Systems}: {Theory}, {Computational} {Methods}, and
  {Functionals},'' in {\em Theoretical and {Computational} {Chemistry}} (J.~M.
  Seminario, ed.), vol.~4 of {\em Recent {Developments} and {Applications} of
  {Modern} {Density} {Functional} {Theory}}, pp.~391--439, Elsevier, Jan. 1996.

\bibitem{casida_molecular_1998}
M.~E. Casida, C.~Jamorski, K.~C. Casida, and D.~R. Salahub, ``Molecular
  excitation energies to high-lying bound states from time-dependent
  density-functional response theory: {Characterization} and correction of the
  time-dependent local density approximation ionization threshold,'' {\em The
  Journal of Chemical Physics}, vol.~108, pp.~4439--4449, Mar. 1998.

\bibitem{stratmann_efficient_1998}
R.~E. Stratmann, G.~E. Scuseria, and M.~J. Frisch, ``An efficient
  implementation of time-dependent density-functional theory for the
  calculation of excitation energies of large molecules,'' {\em The Journal of
  Chemical Physics}, vol.~109, pp.~8218--8224, Nov. 1998.

\bibitem{hirata_time-dependent_1999}
S.~Hirata and M.~Head-Gordon, ``Time-dependent density functional theory within
  the {Tamm}–{Dancoff} approximation,'' {\em Chemical Physics Letters},
  vol.~314, pp.~291--299, Dec. 1999.

\bibitem{furche_adiabatic_2002}
F.~Furche and R.~Ahlrichs, ``Adiabatic time-dependent density functional
  methods for excited state properties,'' {\em The Journal of Chemical
  Physics}, vol.~117, pp.~7433--7447, Oct. 2002.

\bibitem{shao_spinflip_2003}
Y.~Shao, M.~Head-Gordon, and A.~I. Krylov, ``The spin–flip approach within
  time-dependent density functional theory: {Theory} and applications to
  diradicals,'' {\em The Journal of Chemical Physics}, vol.~118,
  pp.~4807--4818, Feb. 2003.

\bibitem{furche_erratum:_2004}
F.~Furche and R.~Ahlrichs, ``Erratum: “{Time}-dependent density functional
  methods for excited state properties” [{J}. {Chem}. {Phys}. 117, 7433
  (2002)],'' {\em The Journal of Chemical Physics}, vol.~121, pp.~12772--12773,
  Dec. 2004.

\bibitem{furche_iii_2005}
F.~Furche and D.~Rappoport, ``{III} - {Density} {Functional} {Methods} for
  {Excited} {States}: {Equilibrium} {Structure} and {Electronic} {Spectra},''
  in {\em Theoretical and {Computational} {Chemistry}} (M.~Olivucci, ed.),
  vol.~16 of {\em Computational {Photochemistry}}, pp.~93--128, Elsevier, Jan.
  2005.

\bibitem{pernal_time-dependent_2007}
K.~Pernal, O.~Gritsenko, and E.~J. Baerends, ``Time-dependent
  density-matrix-functional theory,'' {\em Physical Review A}, vol.~75,
  p.~012506, Jan. 2007.

\bibitem{elliott_excited_2009}
P.~Elliott, F.~Furche, and K.~Burke, ``Excited {States} from {Time}-{Dependent}
  {Density} {Functional} {Theory},'' in {\em Reviews in {Computational}
  {Chemistry}}, pp.~91--165, John Wiley \& Sons, Ltd, 2009.

\bibitem{ipatov_excited-state_2009}
A.~Ipatov, F.~Cordova, L.~J. Doriol, and M.~E. Casida, ``Excited-state
  spin-contamination in time-dependent density-functional theory for molecules
  with open-shell ground states,'' {\em Journal of Molecular Structure:
  THEOCHEM}, vol.~914, pp.~60--73, Nov. 2009.

\bibitem{casida_time-dependent_2009}
M.~E. Casida, ``Time-dependent density-functional theory for molecules and
  molecular solids,'' {\em Journal of Molecular Structure: THEOCHEM}, vol.~914,
  pp.~3--18, Nov. 2009.

\bibitem{ullrich_time-dependent_2011}
C.~A. Ullrich, {\em Time-{Dependent} {Density}-{Functional} {Theory}:
  {Concepts} and {Applications}}.
\newblock Oxford University Press, Dec. 2011.

\bibitem{ferre_density-functional_2016}
N.~Ferré, M.~Filatov, and M.~Huix-Rotllant, eds., {\em Density-{Functional}
  {Methods} for {Excited} {States}}.
\newblock Topics in {Current} {Chemistry}, Springer International Publishing,
  2016.

\bibitem{rocca_ab_2010}
D.~Rocca, D.~Lu, and G.~Galli, ``Ab initio calculations of optical absorption
  spectra: {Solution} of the {Bethe}–{Salpeter} equation within density
  matrix perturbation theory,'' {\em The Journal of Chemical Physics},
  vol.~133, p.~164109, Oct. 2010.

\bibitem{jacquemin_is_2017}
D.~Jacquemin, I.~Duchemin, and X.~Blase, ``Is the {Bethe}–{Salpeter}
  {Formalism} {Accurate} for {Excitation} {Energies}? {Comparisons} with
  {TD}-{DFT}, {CASPT}2, and {EOM}-{CCSD},'' {\em The Journal of Physical
  Chemistry Letters}, vol.~8, pp.~1524--1529, Apr. 2017.

\bibitem{blase_bethesalpeter_2018}
X.~Blase, I.~Duchemin, and D.~Jacquemin, ``The {Bethe}–{Salpeter} equation in
  chemistry: relations with {TD}-{DFT}, applications and challenges,'' {\em
  Chemical Society Reviews}, vol.~47, pp.~1022--1043, Feb. 2018.

\bibitem{gui_accuracy_2018}
X.~Gui, C.~Holzer, and W.~Klopper, ``Accuracy {Assessment} of {GW} {Starting}
  {Points} for {Calculating} {Molecular} {Excitation} {Energies} {Using} the
  {Bethe}–{Salpeter} {Formalism},'' {\em Journal of Chemical Theory and
  Computation}, vol.~14, pp.~2127--2136, Apr. 2018.

\bibitem{krause_implementation_2018}
K.~Krause and W.~Klopper, ``Implementation of the {Bethe}-{Salpeter} equation
  in the {TURBOMOLE} program,'' {\em Journal of Computational Chemistry},
  vol.~38, pp.~383--388, July 2018.

\bibitem{leng_gw_2018}
X.~Leng, F.~Jin, M.~Wei, and Y.~Ma, ``{GW} method and {Bethe}–{Salpeter}
  equation for calculating electronic excitations,'' {\em Wiley
  Interdisciplinary Reviews: Computational Molecular Science}, vol.~6,
  pp.~532--550, July 2018.

\bibitem{foresman_toward_1992}
J.~B. Foresman, M.~Head-Gordon, J.~A. Pople, and M.~J. Frisch, ``Toward a
  systematic molecular orbital theory for excited states,'' {\em The Journal of
  Physical Chemistry}, vol.~96, pp.~135--149, Jan. 1992.

\bibitem{chantzis_is_2013}
A.~Chantzis, A.~D. Laurent, C.~Adamo, and D.~Jacquemin, ``Is the
  {Tamm}-{Dancoff} {Approximation} {Reliable} for the {Calculation} of
  {Absorption} and {Fluorescence} {Band} {Shapes}?,'' {\em J. Chem. Theory
  Comput.}, vol.~9, pp.~4517--4525, Oct. 2013.

\bibitem{mertins_algebraic_1996}
F.~Mertins, J.~Schirmer, and A.~Tarantelli, ``Algebraic propagator approaches
  and intermediate-state representations. {II}. {The} equation-of-motion
  methods for {N} electrons,'' {\em Physical Review A}, vol.~53,
  pp.~2153--2168, Apr. 1996.

\bibitem{schirmer_intermediate_2004}
J.~Schirmer and A.~B. Trofimov, ``Intermediate state representation approach to
  physical properties of electronically excited molecules,'' {\em The Journal
  of Chemical Physics}, vol.~120, pp.~11449--11464, June 2004.

\bibitem{dreuw_algebraic_2015}
A.~Dreuw and M.~Wormit, ``The algebraic diagrammatic construction scheme for
  the polarization propagator for the calculation of excited states,'' {\em
  Wiley Interdisciplinary Reviews: Computational Molecular Science}, vol.~5,
  pp.~82--95, Jan. 2015.

\bibitem{dalgaard_aspects_1983}
E.~Dalgaard and H.~J. Monkhorst, ``Some aspects of the time-dependent
  coupled-cluster approach to dynamic response functions,'' {\em Phys. Rev. A},
  vol.~28, pp.~1217--1222, Sept. 1983.

\bibitem{koch_coupled_1990}
H.~Koch and P.~Joergensen, ``Coupled cluster response functions,'' {\em J.
  Chem. Phys.}, vol.~93, pp.~3333--3344, Sept. 1990.

\bibitem{koch_excitation_1990}
H.~Koch, H.~J.~A. Jensen, P.~Joergensen, and T.~Helgaker, ``Excitation energies
  from the coupled cluster singles and doubles linear response function
  ({CCSDLR}). {Applications} to {Be}, {CH}+, {CO}, and {H}2o,'' {\em J. Chem.
  Phys.}, vol.~93, pp.~3345--3350, Sept. 1990.

\bibitem{christiansen_second-order_1995}
O.~Christiansen, H.~Koch, and P.~Joergensen, ``The second-order approximate
  coupled cluster singles and doubles model {CC}2,'' {\em Chemical Physics
  Letters}, vol.~243, pp.~409--418, Sept. 1995.

\bibitem{hattig_cc2_2000}
C.~Hättig and F.~Weigend, ``{CC}2 excitation energy calculations on large
  molecules using the resolution of the identity approximation,'' {\em J. Chem.
  Phys.}, vol.~113, pp.~5154--5161, Sept. 2000.

\bibitem{hattig_implementation_2002}
C.~Hättig and K.~Hald, ``Implementation of {RI}-{CC}2 triplet excitation
  energies with an application to trans-azobenzene,'' {\em Phys. Chem. Chem.
  Phys.}, vol.~4, pp.~2111--2118, May 2002.

\bibitem{hattig_structure_2005}
C.~Hättig, ``Structure {Optimizations} for {Excited} {States} with
  {Correlated} {Second}-{Order} {Methods}: {CC}2 and {ADC}(2),'' in {\em
  Advances in {Quantum} {Chemistry}} (H.~J.~A. Jensen, ed.), vol.~50 of {\em
  Response {Theory} and {Molecular} {Properties} ({A} {Tribute} to {Jan}
  {Linderberg} and {Poul} {Jørgensen})}, pp.~37--60, Academic Press, Jan.
  2005.

\bibitem{sneskov_excited_2012}
K.~Sneskov and O.~Christiansen, ``Excited state coupled cluster methods,'' {\em
  Wiley Interdisciplinary Reviews: Computational Molecular Science}, vol.~2,
  pp.~566--584, July 2012.

\bibitem{helmich_pair_2013}
B.~Helmich and C.~Hättig, ``A pair natural orbital implementation of the
  coupled cluster model {CC}2 for excitation energies,'' {\em J. Chem. Phys.},
  vol.~139, p.~084114, Aug. 2013.
  
 \bibitem{shao_spinflip_2003}
 Y.~Shao, M.~Head-Gordon, A.I.~Krylov, ``The spin–flip approach within time-dependent density functional theory: Theory and applications to diradicals ," {\em J. Chem. Phys.,} vol.~118, p. 4807, 2003.

\end{thebibliography}

\end{document}